\documentclass[10pt,twoside,preprint2]{aastex}

\usepackage{natbib}
\usepackage{graphicx} 
\usepackage{subfigure}

\usepackage[scaled=1.00]{helvet}

\usepackage[T1]{fontenc}
\usepackage{textcomp}

\usepackage{ragged2e} 

\usepackage{sectsty}
\sectionfont{\normalsize}
\subsectionfont{\normalsize}

\usepackage{fancyhdr}

\usepackage{lscape}
\usepackage{indentfirst}
\usepackage{latexsym}
\usepackage{multirow}
\usepackage{tabls}
\usepackage{wrapfig}
\usepackage{slashbox}
\usepackage{longtable}

\bibpunct{(}{)}{;}{a}{}{,}

\shorttitle{IMF Paper} 
\shortauthors{Popescu et al.}

\begin{document}

\pagestyle{fancy}

\fancyhead{}
\fancyhf{}


\fancyhead[LE,RO]{\bfseries\thepage}

\fancyhead[CE]{{\small {\bf Bogdan Popescu} and {\bf M.M. Hanson} (2013)}}

\fancyhead[CO]{{\footnotesize The Range of Variation of the Mass of the Most Massive Star in Stellar Clusters Derived from 35 Million Monte Carlo Simulations}}


\renewcommand{\thefootnote}{\fnsymbol{footnote}}

\title{
\vspace {1.5cm}
\LARGE{The Range of Variation of the Mass of the Most Massive Star in Stellar Clusters Derived from 35 Million Monte Carlo Simulations}}
\normalsize

\author{{\bf Bogdan Popescu}\altaffilmark{1}\footnote{E-mail: bogdan.popescu@uc.edu} and {\bf M.M. Hanson}\altaffilmark{1}\footnote{E-mail: margaret.hanson@uc.edu} }
\affil{$^{1}$Department of Physics, University of Cincinnati, PO Box 210011, Cincinnati, OH 45221-0011}

\begin{abstract}

A growing fraction of Simple Stellar Population (SSP) models, in an aim to create more realistic simulations capable of including stochastic variation in their outputs, begin their simulations with a distribution of discrete stars following a power-law function of masses.  Careful attention is needed to create a correctly sampled Initial Mass Function (IMF) and in this contribution we provide a solid mathematical method called MASSCLEAN IMF Sampling for doing so. We then use our method to perform {\it 10 million} MASSCLEAN Monte Carlo stellar cluster simulations to determine the most massive star in a mass distribution as a function of the total mass of the cluster. We find a maximum mass range is predicted, not a single maximum mass. This maximum mass range is (a) dependent on the total mass of the cluster and (b) independent of an upper stellar mass limit, $M_{limit}$, for {\it unsaturated} clusters and comes out naturally using our IMF sampling method. We then turn our analysis around, now starting with our new {\it 25 million} simulated cluster database, to constrain the highest mass star from the observed integrated colors of a sample of 40 low-mass LMC stellar clusters of known age and mass. Finally, we present an analytical description of the maximum mass range of the most massive star as a function of the cluster's total mass, and present a new $M_{max}-M_{cluster}$ relation.

\vskip 0.25cm


 {\fontfamily{ptm}\selectfont \textit{The Astrophysical Journal}}{\small, accepted}


\end{abstract}



\section{Introduction}

\renewcommand{\thefootnote}{\arabic{footnote}}
\setcounter{footnote}{0}

It is indeed fortunate that one of the most fundamental of astrophysical distribution functions, the initial mass function (IMF), is described by such a simple equation as a power law.   Yet, this power law, describing the number of stars formed as a function of mass, or the stellar mass spectrum, is central to a broad set of fields in astrophysics.  It is applicable to studies of our solar neighborhood and estimating the number of habitable planets to estimating the mass of the most distant galaxies.  This is possible not just because it is so easily expressed mathematically, but because it's functional form is virtually universal, and thus applicable for studies near and far, past and present. Currently, Salpeter's original paper first reporting the relationship, \citeauthor*{salpeter1955} \citeyearpar{salpeter1955}, garners over 300 citations a year, making it among the most-cited, historical publications in all of astronomy.  \citeauthor*{kroupa2001} \citeyearpar{kroupa2001}, which provides the most accurate present-day values for the exponents used in that power law as a function of mass range, yields another 200 references a year.  The IMF touches, and is deeply fundamental to virtually all of fields astrophysics.  If our use or the analysis of the IMF was in some way wrong or biased, this would have a deeply profound impact in our science.

The apparent simplicity of a power-law equation can be deceiving.  For instance, when attempting to derive the IMF for a population of stars, biases can be introduced in the calculation of the IMF slope for a population of stars, due to the use of constant bin sizes.  In such an analysis, the bins for high mass stars may have very few stars, while bins counting lower mass stars may have tens, hundreds or even thousands of stars.  The bias occurs due to inappropriate weighting of the bins when $\chi^2$ minimization is used to fit the slope.  This was pointed out by many starting over a decade ago (\citeauthor*{kroupa2001} \citeyear{kroupa2001}; \citeauthor*{elmegreen2004} \citeyear{elmegreen2004}; \citeauthor*{jesus2005} \citeyear{jesus2005}). 

What has only more recently been fully recognized, is the bias introduced when the IMF is used to derive a distribution of stars, such as when creating simulations of a stellar cluster or a galaxy.   In some of the earliest models, broadly referred to as simple stellar population (SSP) models (e.g. \citeauthor*{bruzual2003} \citeyear{bruzual2003}) bin size was not a problem because all bins where deamed equal. These models assumed an {\it infinite mass} was available for their stellar distribution. In this kind of analysis, the bins represented the fractional, probabilistic portion compared to the entire stellar distribution. The amount of light or total mass coming from those bins was simply proportional based on this fractional distribution as dictated by stellar evolutionary isochrones. A bias, due to variable weighting was not an issue because discrete stars were not being created. However, such an analysis can not be extended to model the observed properties of low-mass or even mid-mass clusters.  Such models will predict nonphysical values for cluster magnitudes and colors with age, for example calling for a fractional O star. 

If one wishes to simulate a more realistic stellar cluster, down to the tracking of individual stars, and if that cluster is of moderate to low mass (less than $10^4$ $M_{\Sun}$), then assigning stars statistically from a power-law distribution will require real, {\it whole} stars. This also means bin size must be considered, and possible forms of bias need to be identified and addressed.  

In this contribution, we explore the challenges faced with ensuring we create stellar cluster simulations that produce discrete samples of whole stars that fully obey a power-law IMF, but that also assigns masses in such a way that the binning does not lead to any biases in IMF slope or stellar mass range. This later quality will be critical for investigating whether observed clusters in the LMC show evidence for a upper-mass stellar limit $M_{limit}$, or if stellar cluster mass imposes a genuine and biased limit on the most massive star, $M_{max}$, it can form. 

We begin this paper by describing our method to fill the IMF, called MASSCLEAN IMF Sampling. In Section \S 2 we describe the differences between our method and two others: random sampling and {\sc optimal sampling} (\citeauthor*{kroupa2011} \citeyear{kroupa2011}). In Section \S 3 we describe how MASSCLEAN generates the most massive star in the mass distribution of stellar clusters, and compare it with previous work in the field. In Section \S 4 we present the range of variation of the mass of the most massive star as a function of cluster's mass determined from {\it 10 million}  Monte Carlo simulations. We present our new method of deriving the mass of the most massive star using the integrated colors and magnitudes and {\it 25 million} Monte Carlo simulations in Section \S 5. This method is used to estimate the mass of the most massive star for 40 LMC clusters. In Section \S 6 we present an analytical description of the mass range for the most massive star, as well as for the $M_{max} - M_{cluster}$ relation. Concluding remarks are given in Section \S 7.

\section{MASSCLEAN IMF Sampling}

The original theory for the IMF was created, developed, and tested using observational data.  When they were available, the mass of the stars, which form a {\it discrete} distribution, were used to fit a {\it continuous} power law (or multi-power law). The most convenient way for this fitting to be achieved was to use constant mass bins. Another method for measuring the IMF is based on obtaining the K-band luminosity function.  Here, the convenient way of calculating such a function requires setting up constant magnitude bins, which then translate into variable mass bins. So, from the observational point of view, fitting a {\it discrete} distribution to a {\it continuous} power law can be independent of the choice of bins.

However, doing the opposite, filling the {\it continuous} IMF function, to get a {\it discrete} distribution of stars, comes with a whole new set of challenges. The traditional population synthesis models (e.g. \citeauthor*{bruzual2003} \citeyear{bruzual2003}; \citeauthor*{padova2008} \citeyear{padova2008}) were computed in the {\it infinite mass limit}, by assigning a different probability to the stars filling the isochrone. While this works for very massive stellar clusters, for more typical-mass clusters, this corresponds to unphysical, fractional stars, mostly at the upper end of the IMF (e.g. \citeauthor*{paper3} \citeyear{paper3}; \citeauthor*{paper4} \citeyear{paper4}). An alternative method is to randomly populate the IMF.  This method will clearly produce a {\it discrete} distribution. However, like with the binning problem identified in measuring IMF in clusters, this method will also lead to incorrectly populating the cluster, as was shown by \citeauthor*{kroupa2011} \citeyearpar{kroupa2011}, and we will demonstrate here.

For our IMF sampling, we will use our analysis package, MASSCLEAN\footnote{\url{http://www.physics.uc.edu/\textasciitilde popescu/massclean/}\\ {\bf MASS}ive {\bf CL}uster {\bf E}volution and {\bf AN}alysis package is publicly available under GNU General Public License (\copyright 2007-2013 Bogdan Popescu and Margaret Hanson).} (\citeauthor*{paper1} \citeyear{paper1}).  A thorough description of the code is available from our earlier papers (\citeauthor*{paper1} \citeyear{paper1}, \citeyear{paper2}, \citeyear{paper3}; \citeauthor*{paper4} \citeyear{paper4}), but perhaps most unique is that it allows for a realistic representation of the stochastic fluctuations that occur in real clusters, which is increasingly important as the mass of the cluster decreases. We will provide an outline of the critical aspects of the simulation that apply to this investigation below.

The mass distribution of stars in stellar clusters is described by the IMF, so the number of stars formed in the $M\pm\mathrm{d}M$ range is:

   \begin{equation} \label{eq:1}
    \mathrm{d}N=\xi(M) \mathrm{d}M
   \end{equation}

\noindent
where 
 $\xi(M)$ is the Kroupa-Salpeter IMF (\citeauthor*{kroupa2001} \citeyear{kroupa2001}, \citeyear{kroupa2002}; \citeauthor*{kroupa2011} \citeyear{kroupa2011}; \citeauthor*{salpeter1955} \citeyear{salpeter1955}):
\begin{equation}  \label{eq:2}
    \xi (M) = k \left \{ 
    \begin{array}{lcc} 
    \left (  \frac{M}{m_{1}} \right )^{-\alpha_{1}} &,& m_{0}<M \leq m_{1} \\
    \left (  \frac{M}{m_{1}} \right )^{-\alpha_{2}} &,& m_{1}<M \leq m_{2} \\
    \left (  \frac{m_{2}}{m_{1}} \right )^{-\alpha_{2}} \left (  \frac{M}{m_{2}} \right )^{-\alpha_{3}} &,& m_{2}<M \leq m_{3} 
    \end{array}
    \right.  
   \end{equation}
   with mass expressed in $M_{\sun}$ units. For this work we used:
\begin{equation} \label{eq:3}
   \begin{array}{lcc} 
   \alpha_{1}=+0.30 ,& 0.01 \leq M/M_{\sun}<0.08 \\
   \alpha_{2}=+1.30 ,& 0.08 \leq M/M_{\sun}<0.50 \\
   \alpha_{3}=+2.35 ,& 0.50 \leq M/M_{\sun}<m_{3} 
   \end{array}
   \end{equation}
and $m_{3}=M_{cluster}$ (the total mass of the cluster) or $m_{3}=M_{limit}$ (for an IMF with upper mass cutoff (e.g. \citeauthor*{oey2005} \citeyear{oey2005}; \citeauthor*{kroupa2011} \citeyear{kroupa2011})). Note that for $\alpha_{1}=\alpha_{2}=\alpha_{3}=2.35$, $\xi(M)$ becomes the \citeauthor*{salpeter1955} \citeyearpar{salpeter1955} IMF.

Using $\xi (M)/k = \xi_{i} (M)$ (with $i=1,2,3$ respectively), the IMF could be simplified to:

\begin{equation} \label{eq:4}
   \xi (M) = k \: \xi_{i} (M)
   \end{equation}
Then the total mass of the cluster can be written :

\begin{equation} \label{eq:5}
    M_{cluster}= \int_{0}^{N_{max}} M(N) \mathrm{d}N 
   \end{equation}
   
\begin{equation} \label{eq:6}
    M_{cluster}=  \int_{m_{0}}^{m_{3}} M \frac{\mathrm{d}N }{\mathrm{d}M } \mathrm{d}M = \int_{m_{0}}^{m_{3}} \xi (M) M \mathrm{d}M 
   \end{equation}   
   
\begin{equation} \label{eq:7}
    M_{cluster}=\sum_{i=1}^{3} \left( k \int_{m_{i-1}}^{m_{i}} \xi_{i} (M) M \mathrm{d}M  \right)
\end{equation}   
   
The normalization constant :   
\begin{equation} \label{eq:8}
   k =\frac{ M_{cluster}} {\sum_{i=1}^{3} \left(    \int_{m_{i-1}}^{m_{i}} \xi_{i} (M) M \mathrm{d}M  \right)}
\end{equation}   

From the equations (\ref{eq:1}), (\ref{eq:2}), (\ref{eq:4}) and (\ref{eq:8}) we get an equation that describes each bin: 
\begin{equation} \label{eq:9}
   N_{i}(M, \Delta M) =\frac{ M_{cluster} \int_{M-\Delta M}^{M+\Delta M} \xi_{i}(M) \mathrm{d}M} {\sum_{i=1}^{3} \left( \int_{m_{i-1}}^{m_{i}} \xi_{i} (M) M \mathrm{d}M  \right)}
\end{equation}

Let's note that the sum of integrals is a constant:

\begin{equation} \label{eq:10}
   C =\sum_{i=1}^{3} \left( \int_{m_{i-1}}^{m_{i}} \xi_{i} (M) M \mathrm{d}M  \right)
\end{equation}
so equation (\ref{eq:9}) could be written:

\begin{equation} \label{eq:11}
   N_{i}(M, \Delta M) =\frac{ M_{cluster}} {C} \int_{M-\Delta M}^{M+\Delta M} \xi_{i}(M) \mathrm{d}M 
\end{equation}

Equation (\ref{eq:11}) could be used to compute the number of stars in any mass interval, as described before in \citeauthor*{paper1} \citeyearpar{paper1}. It could also be used to compute the probabilities or the multiplication factors corresponding to isochrone stars used in the traditional SSP models computed in the {\it infinite mass limit} (e.g. \citeauthor*{padova2008} \citeyear{padova2008}; \citeauthor*{padova2010} \citeyear{padova2010}).

However, if we wish to create a {\it discrete} distribution of stars in a cluster, this can not include a {\it fractional} number of stars. Thus, the challenge is to compute the $M_{j-1}$ and $M_{j}$ mass limits such that $N_{i}(M_{j-1},M_{j})$ will always give an integer value.  In \citeauthor*{kroupa2011} \citeyearpar{kroupa2011}, they describe a similarly derived set of equations and forced each bin to contain precisely 10 stars.  They call this {\sc optimal sampling}, where the IMF is perfectly sampled without any gaps in the distribution and the stellar masses are ideally spaced.  They suggest that this {\sc optimal sampling} is a more realistic approach to populating the IMF, versus the more traditionally used random sampling of the IMF (\citeauthor*{kroupa2011} \citeyear{kroupa2011}). 

Using Equation (\ref{eq:11}), MASSCLEAN also computes an integer value of stars per variable mass bin.  However, for our simulations, we define each bin to include exactly one star.  In other words, we set $N_{i}(M_{j-1},M_{j})= 1$.  We will call this method {\bf M}ASSCLEAN {\bf IMF} {\bf S}ampling (MIMFS).  The expanded utility of this choice over the \citeauthor*{kroupa2011} \citeyearpar{kroupa2011} {\sc optimal sampling} method will become apparent soon.  

Let's switch to the notation $M_{j}=M-\Delta M$ and $M_{j-1}=M+\Delta M$, with $j=1$ to $N_{max}$. When:
\begin{equation} \label{eq:12}
   N_{i}(M_{j}, M_{j-1}) =\frac{ M_{cluster}} {C} \int_{M_{j}}^{M_{j-1}} \xi_{i}(M) \mathrm{d}M = 1
\end{equation}
the choice of $j$ shows that the $(M_{j},M_{j-1})$ bin contains the $j$th most massive star.  In this notation, the most massive star in the cluster will be found in the $(M_{1},M_{0})$ bin.

For another set of constants, let's use the notation: $\gamma_{i}=\xi_{i}(M)/M^{-\alpha_{i}}$, (with $i=1,2,3$ respectively). From the Equation (\ref{eq:12}) we can compute $M_{j}$:

\begin{equation} \label{eq:13}
   M_{j} =\left( M_{j-1} -\frac{C}{M_{cluster}} \frac{1-\alpha_{i}}{\gamma_{i}} \right)^{\frac{1}{1-\alpha_{i}}   }
\end{equation}

Consequently, the interval $(M_{j-1+N}, M_{j-1})$ will contain $N$ stars, with:

\begin{equation} \label{eq:14}
   M_{j-1+N} =\left( M_{j-1} -\frac{C N}{M_{cluster}} \frac{1-\alpha_{i}}{\gamma_{i}} \right)^{\frac{1}{1-\alpha_{i}}   }
\end{equation}

\begin{figure}[htp]
\centering
\subfigure[]{\includegraphics[angle=270,width=0.48\textwidth]{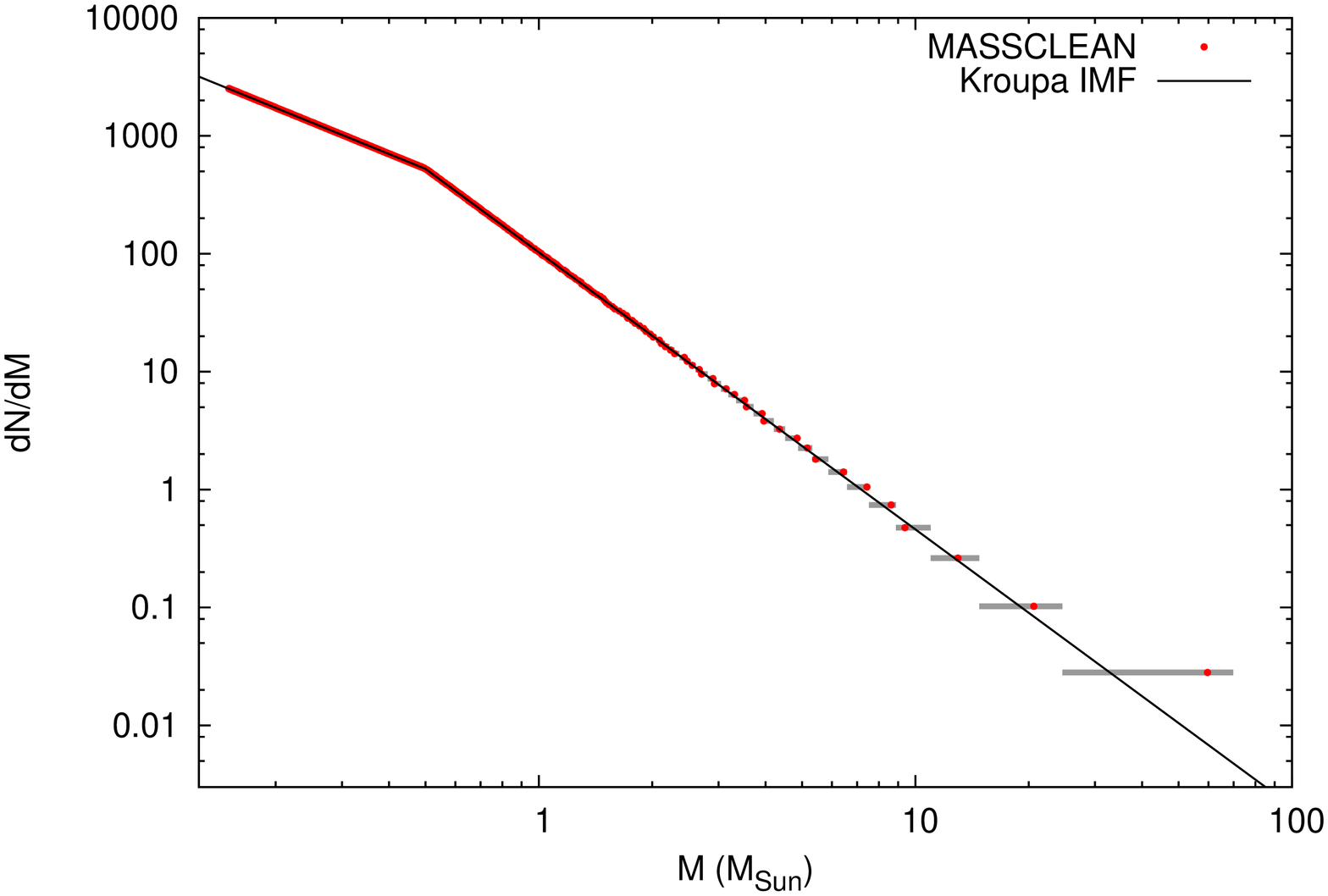}} 
\subfigure[]{\includegraphics[angle=270,width=0.48\textwidth]{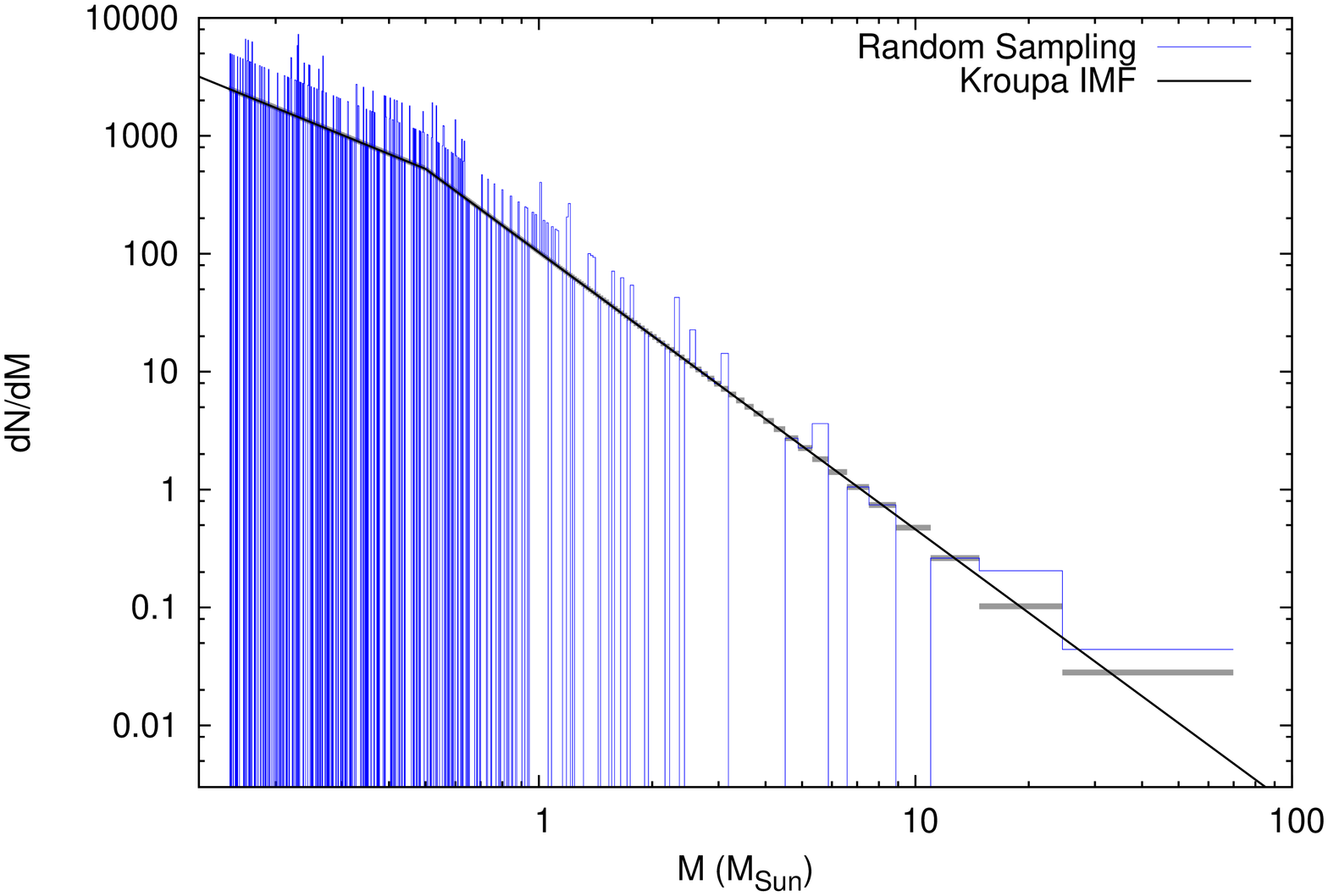}} 
\subfigure[]{\includegraphics[angle=270,width=0.48\textwidth]{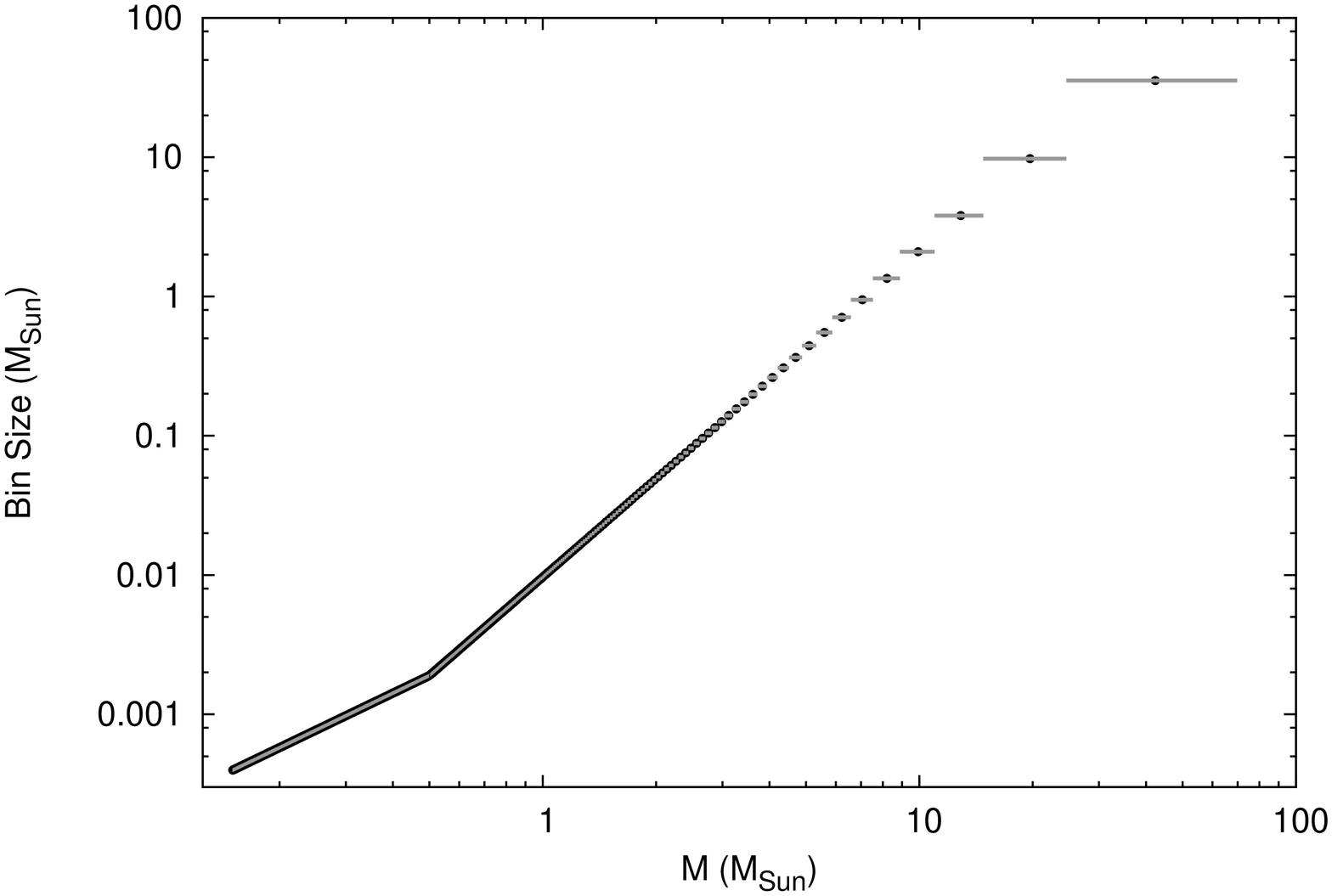}} 
\caption[]{\small The Stellar IMF.   (a) The diagonal line represents the Kroupa-Salpeter IMF as given in Equation (\ref{eq:2}).  The segmented horizontal lines represent the variable mass bins created using our MASSCLEAN IMF Sampling (MIMFS) method, and assuming a total cluster mass of $500$ $M_{\Sun}$.  Discrete stars are given by the red dots, with one per MIMFS mass bin.  (b) The same as (a), only overlaid in blue is the result of a stellar distribution created using a randomly populated IMF, leaving numerous gaps.  (c) Showing the mass-dependent bins of our MIMFS method.     \normalsize}\label{fig:paper5-01}
\end{figure}

In Figure \ref{fig:paper5-01} (a) we show an example demonstrating the MIMFS method for distributing stars for a stellar cluster with $M_{cluster}=500$ $M_{\Sun}$. The \citeauthor*{kroupa2001} \citeyearpar{kroupa2001} IMF is presented as the diagonal black line, and the $(M_{j},M_{j-1})$ MASSCLEAN variable-mass bins are displayed as horizontal gray lines. The mass values for all of the stars created as part of this cluster from a sample MASSCLEAN run are shown as the red dots. This sample mass distribution is consistent with the continuous IMF power-law.  However, it also includes very real fluctuations, as the stellar mass, $M$, is allowed to fall anywhere in the $(M_{j},M_{j-1})$ bin. Although the fluctuations of these red dots from the black line representing the \citeauthor*{kroupa2001} \citeyearpar{kroupa2001} IMF value may appear small, they can generate a large dispersion in the integrated magnitudes and colors of simulated clusters, and are consistent with available observational data (\citeauthor*{paper2} \citeyear{paper2}, \citeyear{paper3}; \citeauthor*{paper4} \citeyear{paper4}).  This is because the fluctuations (mass range allowed) within each bin is proportional to the bin mass, with the highest mass fluctuations occurring among the most massive stars in the cluster.  The MIMFS method correctly simulates the largest variation in integrated magnitude and color to be seen among the low-mass clusters as they have relatively few, very large bins at the high mass end. 

In Figure \ref{fig:paper5-01} (b) we present an example mass distribution generated using the traditional random sampling of the IMF with the same $M_{cluster}=500$ $M_{\Sun}$ stellar cluster. To compare to our MIMFS method, we also plot the same greyscale lines representing MASSCLEAN bins as presented in Figure  \ref{fig:paper5-01} (a).  As already pointed out by \citeauthor*{kroupa2011} \citeyearpar{kroupa2011}, the traditional method of random sampling shows large variations from the IMF, with many large gaps.  As discussed above, these unnatural variations will not disappear by simply using a different choice of bin size.  While random sampling will work in the realm of deriving {\sl relative} fractional stars in the limit of an {\it infinite mass} distribution, it can not be used to populate the IMF properly when a discrete stellar distribution is needed, such as when simulating low-mass clusters (\citeauthor*{jesus2009} \citeyear{jesus2009}).

Note that in both the \citeauthor*{kroupa2011} \citeyearpar{kroupa2011} {\sc optimal sampling} and our own MIMFS methods, bin size will be mass-dependent.  The size of the bins as a function of mass is presented in Figure \ref{fig:paper5-01} (c). The bin size $\Delta M_{bin} = M_{j-1} - M_{j}$ is expressed on the vertical axes as black dots.   The limits of the bins are represented along the horizontal axis with the gray lines. The plot shows that $log(\Delta M_{bin}) \propto log(M)$.  What's more, as described by the Equation (\ref{eq:13}), the bin size also depends on the mass of the cluster, $M_{cluster}$.   This is an obvious result, if one recalls we are forcing each bin to hold just one star.  More massive clusters will have proportionally more bins over the same stellar mass range. 

\begin{figure}[htp]
\centering
\subfigure[]{\includegraphics[angle=270,width=0.48\textwidth]{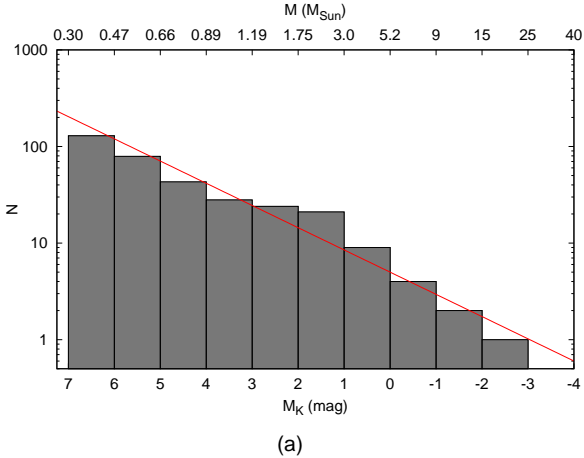}} 
\caption[]{\small  Demonstrates a K-band luminosity function, derived using MIMFS for a $M=500$ $M_{\Sun}$ cluster, 10$^6$ years old.  \normalsize}\label{fig:paper5-02}
\end{figure}

Figure \ref{fig:paper5-02} is an example of a $K$ band luminosity function, as derived using our MIMFS method. The sample cluster presented in Figure \ref{fig:paper5-02} was simulated with MASSCLEAN and is 1 million yrs old, with $M=500$ $M_{\Sun}$. The slope in the histogram depends on the mass-to-light ratio and on the $\alpha_{3}$ value.  It can also be used to constrain the stellar IMF.  The constant magnitude bins in the histogram (labeled in the lower axis) correspond to the logarithmic, age-dependent bins in mass (labeled in the upper axis).

\section{The Most Massive Star in a Stellar Cluster}

There are strong similarities between Kroupa's {\sc optimal sampling} (\citeauthor*{kroupa2011} \citeyear{kroupa2011}) and our MIMFS method for discretely populating stellar clusters: the bins are mass-dependent and the number and bin sizes are dependent on the cluster mass. This leads to a stellar mass distribution that truly obeys the IMF.  But there is one significant difference in the two methods. By forcing our bins in the MIMFS method to contain no more than one star, we are able to go one step further to make analytical predictions for the mass of the most massive star in a cluster.

Following the formalism and notation presented in section \S 2, the most massive star in the mass distribution could be described using Equation (\ref{eq:12}):

\begin{equation} \label{eq:15}
  \frac{M_{cluster}}{C} \int_{M_{1}}^{M_{0}} \xi_{3} (M) \mathrm{d}M = N_{3}(M_{1},M_{0}) = 1  
\end{equation} 

\begin{equation} \label{eq:16}
   M_{cluster} \frac{\gamma_{3}}{C}\int_{M_{1}}^{M_{0}} M^{-\alpha_{3}} \mathrm{d}M = N_{3}(M_{1},M_{0}) = 1  
\end{equation}   

For the upper limit we will start with three choices: 
$M_{0}=\infty$ (e.g. \citeauthor*{elmegreen2000} \citeyear{elmegreen2000}, for the convenience of the computation and description); $M_{0}=M_{cluster}$ (since obviously the most massive star could not have a mass bigger than the entire cluster mass); and $M_{0}=M_{limit}$ (the maximum stellar mass, e.g. \citeauthor*{oey2005} \citeyear{oey2005}).

When $M_{0}=\infty$, $M_{1}$ can be determined from Equation (\ref{eq:16}):

\begin{equation} \label{eq:17}
   M_{1}= \left( \frac{\gamma_{3}}{C} \frac{1}{\alpha_{3}-1} \right)^{\frac{1}{\alpha_{3}-1}} M_{cluster}^{\frac{1}{\alpha_{3}-1}}
\end{equation} 

Adding numbers, this simplifies to:

\begin{equation} \label{eq:18}
   M_{1}= 0.2375 \phn M_{cluster}^{\frac{1}{1.35}}
\end{equation} 

This is virtually identical to the relation found by \citeauthor*{elmegreen2000} \citeyearpar{elmegreen2000}, who used a similar formalism, but a different IMF:

\begin{equation} \label{eq:19}
   M_{max}= 100  \left( \frac{M_{cluster}}{3 \times 10^{3}}    \right)^{\frac{1}{1.35}}
\end{equation} 

Both $M_{1}$ and $M_{max}$ given by Equations (\ref{eq:18}) and (\ref{eq:19}), respectively, are proportional to $M_{cluster}^{\frac{1}{1.35}}$. 

When the upper mass limit is changed to $M_{0}=M_{cluster}$, from Equation (\ref{eq:16}) we instead get:

\begin{equation} \label{eq:20}
   M_{1}= \left(M_{cluster}^{1-\alpha_{3}} -\frac{C}{M_{cluster}} \frac{1-\alpha_{3}}{\gamma_{3}}      \right)^{\frac{1}{1-\alpha_{3}}}
\end{equation} 

Again, adding numbers, this simplifies to:

\begin{equation} \label{eq:21}
   M_{1}= \left( M_{cluster}^{-1.35} -\frac{6.9653}{M_{cluster}}  \right)^{-\frac{1}{1.35}}
\end{equation} 

\begin{figure}[htp]
\centering
\subfigure[]{\includegraphics[angle=270,width=0.48\textwidth]{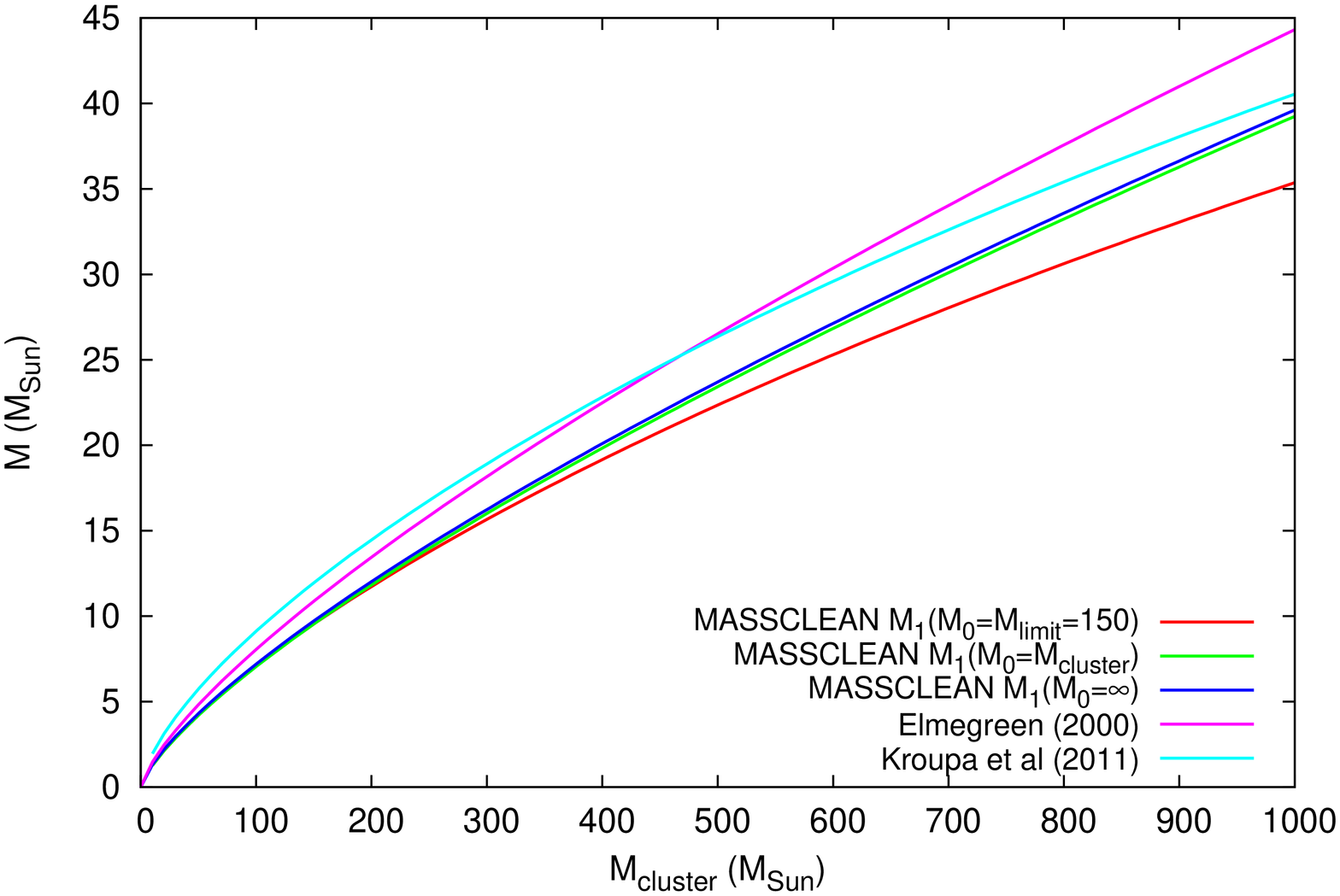}} 
\subfigure[]{\includegraphics[angle=270,width=0.48\textwidth]{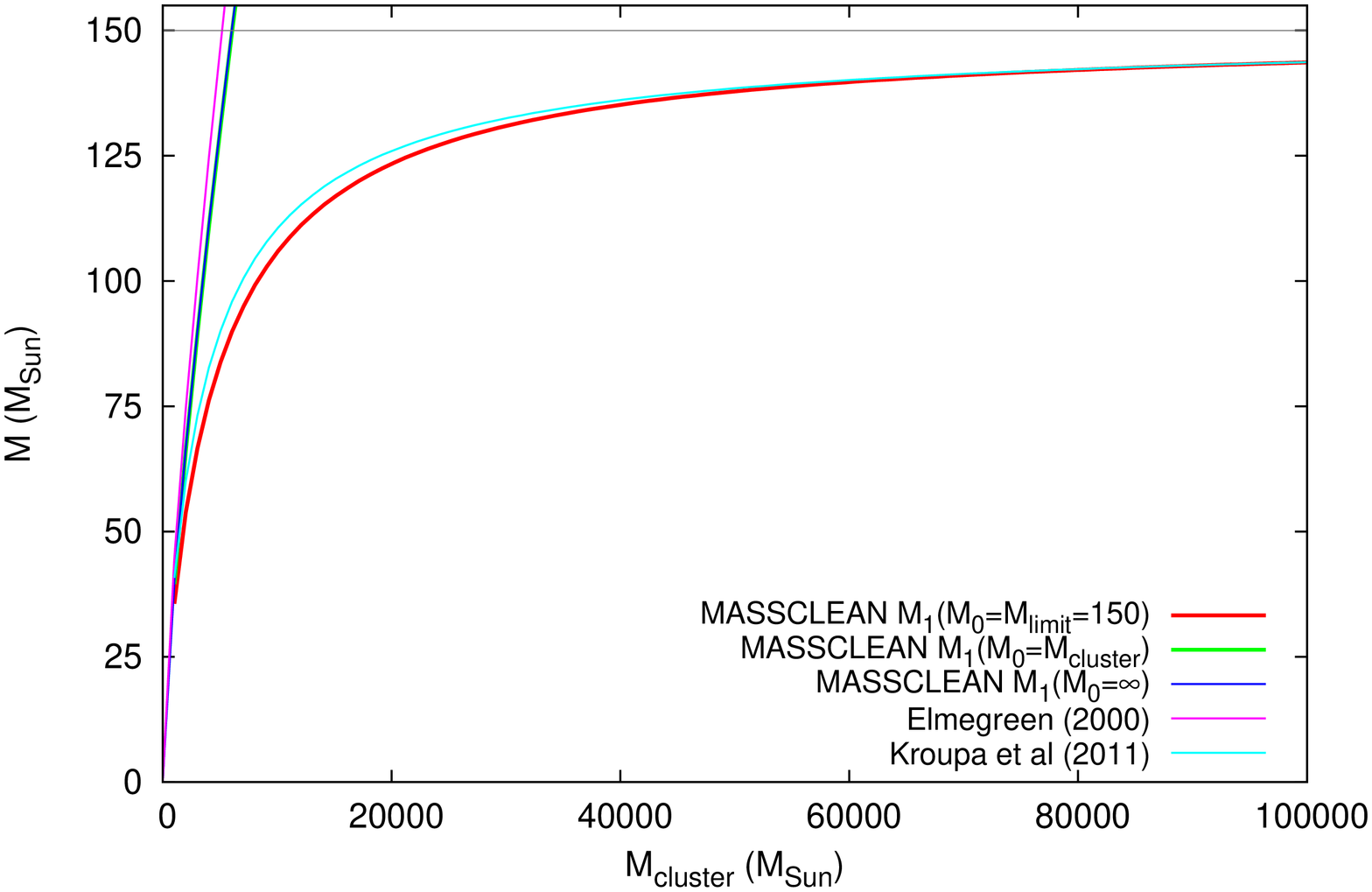}} 
\subfigure[]{\includegraphics[angle=270,width=0.48\textwidth]{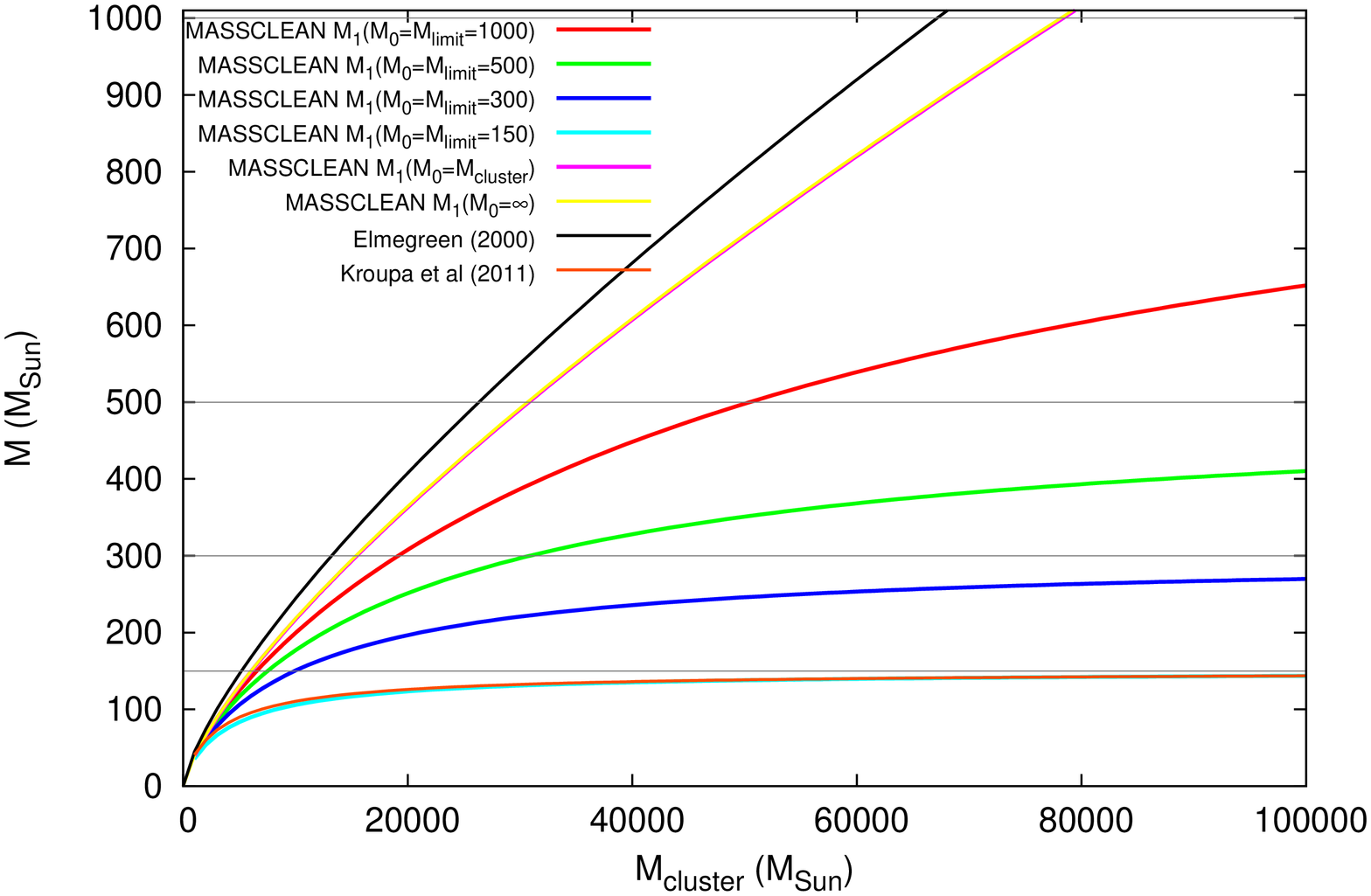}} 
\caption[]{\small Graphical representations of the most massive star possible in a cluster versus the cluster total mass, using the variant expressions discussed in \S 3.  (a) For modest cluster masses, $M_{cluster} < 1,000$ $M_{\Sun}$, the differing limit chosen for the most massive star has a somewhat minor effect on the predicted most massive star observed in the host cluster.  (b) Once the cluster mass becomes fairly large, and beyond a mass of $10,000$ $M_{\Sun}$, an increasingly large divergence is seen in the predicted maximum mass and between expressions allowing an infinite stellar mass and those limiting the stellar mass.   (c)  A closer look at the predicted maximum stellar mass, based on differing upper mass limits, from $M_{\max} =$ $150$ through $1,000$ $M_{\Sun}$.  \normalsize}\label{fig:paper5-03}
\end{figure}

Although it seams reasonable to use $M_{cluster}$ as an upper mass limit instead of $\infty$, the difference between Equation (\ref{eq:18}) and Equation (\ref{eq:21}) is indeed very small.  Figure \ref{fig:paper5-03} (a) shows the shape of these two functions.  They are virtually identical, with $M_{0} = \infty$ predicting only a slightly more massive star formed in the corresponding cluster.

Changing the upper mass limit to a specific upper mass maximum value, $M_{limit}$ in Equation (\ref{eq:16}) gives:

\begin{equation} \label{eq:22}
   M_{1}= \left(M_{limit}^{1-\alpha_{3}} -\frac{C}{M_{cluster}} \frac{1-\alpha_{3}}{\gamma_{3}}      \right)^{\frac{1}{1-\alpha_{3}}}
\end{equation} 

After simplifying the numbers, this leads to:

\begin{equation} \label{eq:23}
   M_{1}= \left( M_{limit}^{-1.35} -\frac{6.9653}{M_{cluster}}  \right)^{-\frac{1}{1.35}}
\end{equation} 

This expression is close to \citeauthor*{kroupa2011} \citeyearpar{kroupa2011}, who found:
\begin{eqnarray} \label{eq:24}
   \log(M_{max}) =  2.56\log(M_{cluster})[3.82^{9.17}+ \nonumber \\
                  + \log(M_{cluster})^{9.17}]^{-\frac{1}{9.17}}-0.38 
\end{eqnarray} 
\noindent
and applied a canonical upper mass limit of $M_{limit}=150 M_{\Sun}$.

The $M_{1}$ variation described by Equations (\ref{eq:18}), (\ref{eq:20}), and (\ref{eq:21}), along with $M_{max}$ variation from \citeauthor*{elmegreen2000} \citeyearpar{elmegreen2000} and \citeauthor*{kroupa2011} \citeyearpar{kroupa2011}, given by Equations (\ref{eq:19}) and (\ref{eq:24}) respectively, are presented in Figure \ref{fig:paper5-03}. In Figure \ref{fig:paper5-03} (a), the maximum mass of the cluster is $1,000$ $M_{\Sun}$. In this mass range there is a fairly small difference between all of the mentioned variations for setting the upper mass limit. In the case of $M_{1}$, as we noted earlier, there is a negligible difference when switching from $M_{0}=\infty$ to $M_{0}=M_{cluster}$. 

In Figure \ref{fig:paper5-03} (b) the mass range is now expanded to $100,000$ $M_{\Sun}$. In this range it is now easy to see that $M_{max}$ from \citeauthor*{elmegreen2000} \citeyearpar{elmegreen2000} and $M_{1}$ as given by our Equations (\ref{eq:18}) and (\ref{eq:20}) continue to agree, yet quickly diverge from the other functions.  As expected, \citeauthor*{kroupa2011} \citeyearpar{kroupa2011} and Equation (\ref{eq:23}), with $M_{limit}=150 M_{\Sun}$ remain in good agreement over this entire range.

In Figure \ref{fig:paper5-03} (c), the $y$-axis range is expanded to show the variation found in deriving $M_{1}$, given by the Equation (\ref{eq:23}), for differing values for the upper mass maximum value, using limiting values of $M_{lim}=150$ $M_{\Sun}$, $300$ $M_{\Sun}$, $500$ $M_{\Sun}$, and $1,000$ $M_{\Sun}$.

Both \citeauthor*{elmegreen2000} \citeyearpar{elmegreen2000} and \citeauthor*{kroupa2011} \citeyearpar{kroupa2011} use the notation $M_{max}$ in Equations (\ref{eq:19}) and (\ref{eq:24}), respectively. Using a similar formalism, we introduced $M_{1}$ in Equations (\ref{eq:18}), (\ref{eq:21}), and (\ref{eq:23}). Obviously, the most massive star, with the mass $M_{max}$, is most likely to be in the $(M_{1},M_{0})$ interval. $M_{1}$ is only a measure of the {\it lower mass limit} of the most massive star.  $M_{max}$, the most massive star, could have a mass as high as $M_{0}$. When the mass of the cluster is high enough $M_{0}$, the highest mass possible for the star will hit the limit of the upper mass maximum value, $M_{limit}$.  These clusters are referred to as {\it saturated} (\citeauthor*{kroupa2011} \citeyear{kroupa2011}). For lower mass clusters, when $M_{limit}$ is not reached, $M_{0}$ will have a range of variation depending on the total mass of the cluster, $M_{cluster}$, as discussed in the next sections. 

How does the MIMFS method differ from the {\sc optimal sampling} of \citeauthor*{kroupa2011} \citeyearpar{kroupa2011}? In both cases the IMF is filled properly (Figure \ref{fig:paper5-01} (a)), and the bins, for {\it one star} (MIMFS) or  {\it some constant integer number of stars} ({\sc optimal sampling}), are both mass-dependent and cluster-mass-dependent (Figure \ref{fig:paper5-01} (a) and (c)).  However, the treatment of the most massive star is different.  With {\sc optimal sampling}, a limit on the mass of the most massive star is used, a single valued $M_{max}$$-$$M_{cluster}$ relation, computed for $M_{limit}=150$ $M_{\Sun}$ ({\it canonical limit}, Equation (\ref{eq:24})).   

The MIMFS method does not use a predefined, canonical, maximum mass limit for the most massive star, even for those clusters with too low a mass to expect a star to be found above the physical limit ({\it unsaturated clusters}). Instead, each star mass, $M_{k}$, is simply assigned randomly in the $(M_{j},M_{j-1})$ interval (Equations (\ref{eq:12}) and (\ref{eq:13})).  The upper limit of the mass of the most massive star in the cluster will be given by:

\begin{equation} \label{eq:25}
   M_{0}= M_{cluster} - \sum_{k=1}^{N_{max}-1} M_{k}
\end{equation} 

$M_{0}$ is the maximum available mass for the most massive star, and it depends on the entire distribution of stars. Our $10$ $million$ Monte Carlo simulations show that $M_{max}$ could be significantly {\it higher} than the {\it canonical limit} used by \citeauthor*{kroupa2011} \citeyearpar{kroupa2011}, which is well approximated by $M_{1}$ (Equation (\ref{eq:23})). 

However, due to the stochastic fluctuations in each mass bin, $M_{max}$ could also be {\it lower} than $M_{1}$. This is because the {\it remaining mass} for the most massive star (Equation (\ref{eq:25})) could be lower than $M_{1}$ for {\it unsaturated} clusters.  This is related to the way the bins are filled with the MIMFS method.   Nearly all of the bins, particularly for a small cluster, will populate the distribution with very low-mass stars.  That any bin would be assigned a star is equally likely (the size of the bin has been specifically created to match the form of the IMF).  In {\it unsaturated} clusters, if the highest mass bin is not filled until after most of the low-mass stars bins have been populated, it may not be possible to fill that bin because the mass left for stars is less than $M_{1}$.  In this situation, the Equation (\ref{eq:14}) shows that the $(M_{2},M_{0}=M_{limit})$ interval contains $N=2$ most massive stars:

\begin{equation} \label{eq:26}
   M_{2} =\left( M_{limit} -\frac{2 C}{M_{cluster}} \frac{1-\alpha_{3}}{\gamma_{3}} \right)^{\frac{1}{1-\alpha_{3}}   }
\end{equation}

Although this is a rare event, the two most massive stars could instead be located in the combined $(M_{2},M_{limit})$ bin. So, $M_{2}$ could also be used as a measure of the lower limit of $M_{max}$ for low-mass, {\it unsaturated} clusters.

\section{Mass Range of the Most Massive Star as a Function of Cluster's Mass}

It is reasonable to assume there is a genuine, maximum mass of a star, $M_{limit}$, based on physical processes allowing for a structurally stable star (\citeauthor*{stothers1992} \citeyear{stothers1992}; \citeauthor*{baraffe2001} \citeyear{baraffe2001}; \citeauthor*{massey2011} \citeyear{massey2011}). Presently, models of stellar structure are not able to provide a strong constraint on this limit.  However, using star counts and statistical arguments, a {\it canonical} value of $M_{limit}$ $\sim 150$ $M_{\Sun}$ has been claimed (e.g.\ \citeauthor*{weidner2004} \citeyear{weidner2004}; \citeauthor*{oey2005} \citeyear{oey2005}).  This limit has also been argued based on observations made of the most massive stars in the very young, high-mass clusters of {\it R136} in the Large Magellanic Clouds (\citeauthor*{selman1999} \citeyear{selman1999}) and the {\it Arches} cluster near the center of the Milky Way (\citeauthor*{figer2005} \citeyear{figer2005}).  

However, there has recently emerged some refinements concerning the accepted masses of the most massive stars in these clusters. \citeauthor*{crowther2010} \citeyearpar{crowther2010} has applied a modified spectral analysis to several stars in {\it NGC 3603}, {\it Arches}, and {\it R136} and conclude the stellar mass for some of these stars exceeds the {\it canonical} limit of $150$ $M_{\Sun}$.  Obtaining masses of high mass stars from atmospheric analysis is a tricky business, particularly at extremely high mass.  Moreover, there is considerable evidence to suggest stellar masses, derived using spectroscopic analysis, may be underestimating the mass of high-mass stars, the so-called mass discrepancy problem first pointed out by \citeauthor*{herrero1992} \citeyearpar{herrero1992}.  Regretfully, there are few high-mass binaries with extreme masses of $120$ or even $150$ $M_{\Sun}$) to help calibrate these analyses (though note the recent identification that {\it R144} in {\it 30 Doradus} is a binary of combined mass nearing $400$ $M_{\Sun}$, e.g. \citeauthor*{sana2013} \citeyear{sana2013}). 

We have used MASSCLEAN to perform $10$ $million$ Monte Carlo simulations in order to determine $M_{max}$ as a function of $M_{cluster}$. We investigate the properties of stellar clusters where the stellar mass limit $M_{limit}$ is set to $150, 300, 500$ and $1,000$ $M_{\Sun}$, and as part of a cluster of mass between $10$ and $100,000$ $M_{\Sun}$. The mass distribution was computed using the MIMFS algorithm, described in \S 2 and \S 3, and thus the IMF is always filled properly without any gaps. However, due to natural, stochastic fluctuations in the IMF, the mass of the most massive star, $M_{max}$, is not single valued.  Instead, because of using the MIMFS method, we derive the expected {\it mass range} for the most massive star in a cluster.

\begin{figure*}[htp]
\centering
\subfigure[]{\includegraphics[angle=270,width=0.48\textwidth]{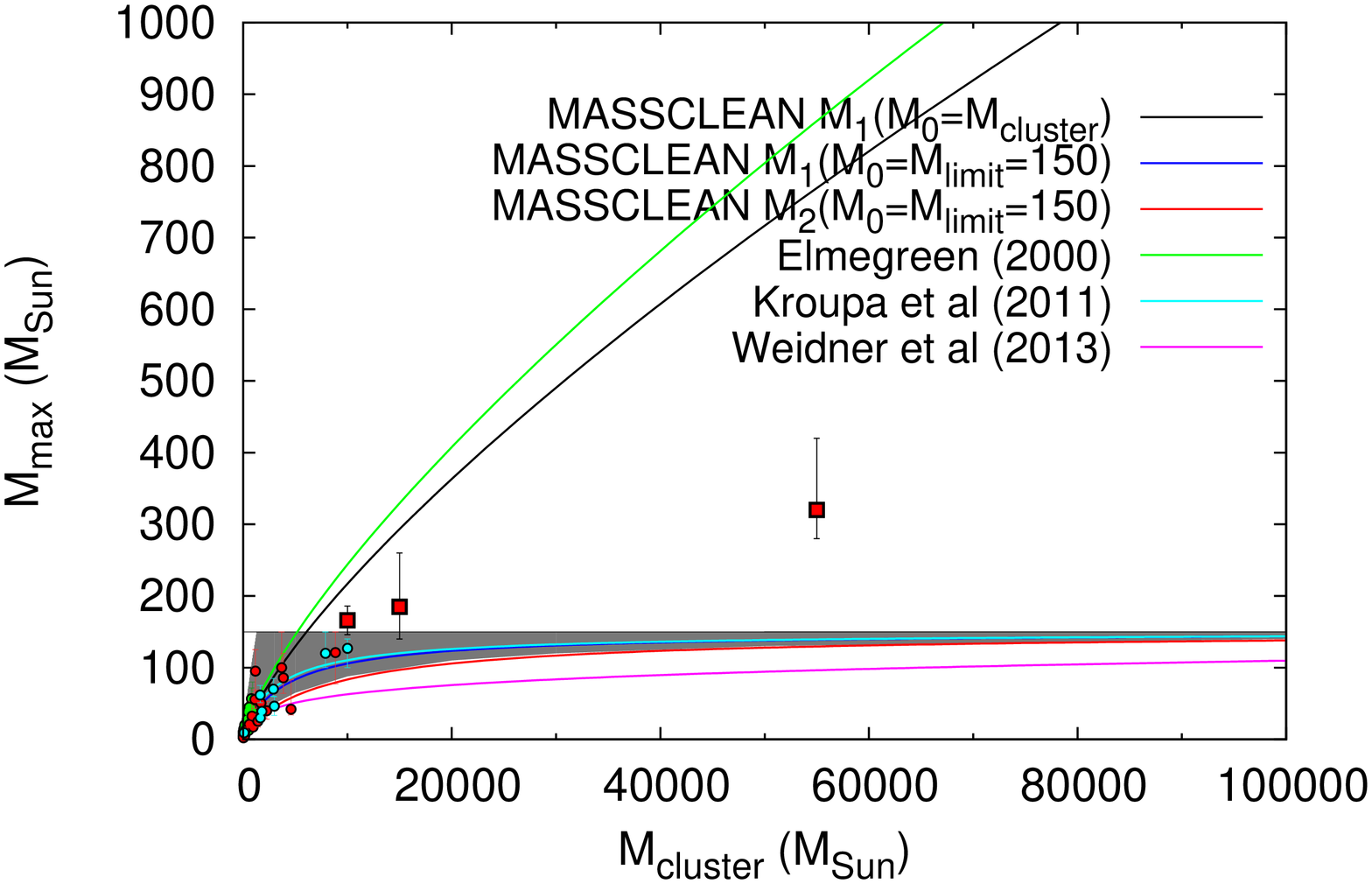}} 
\subfigure[]{\includegraphics[angle=270,width=0.48\textwidth]{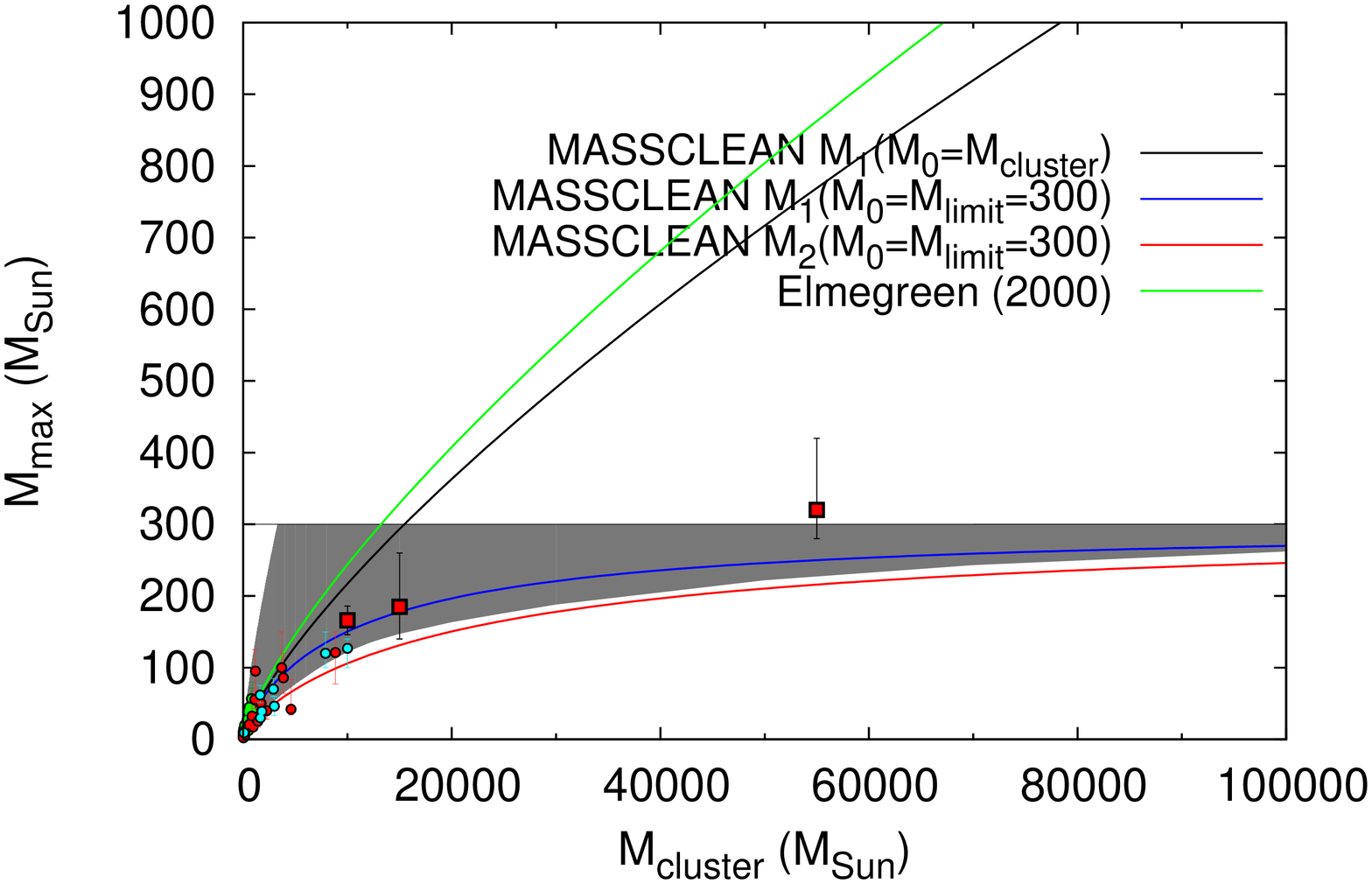}} 
\subfigure[]{\includegraphics[angle=270,width=0.48\textwidth]{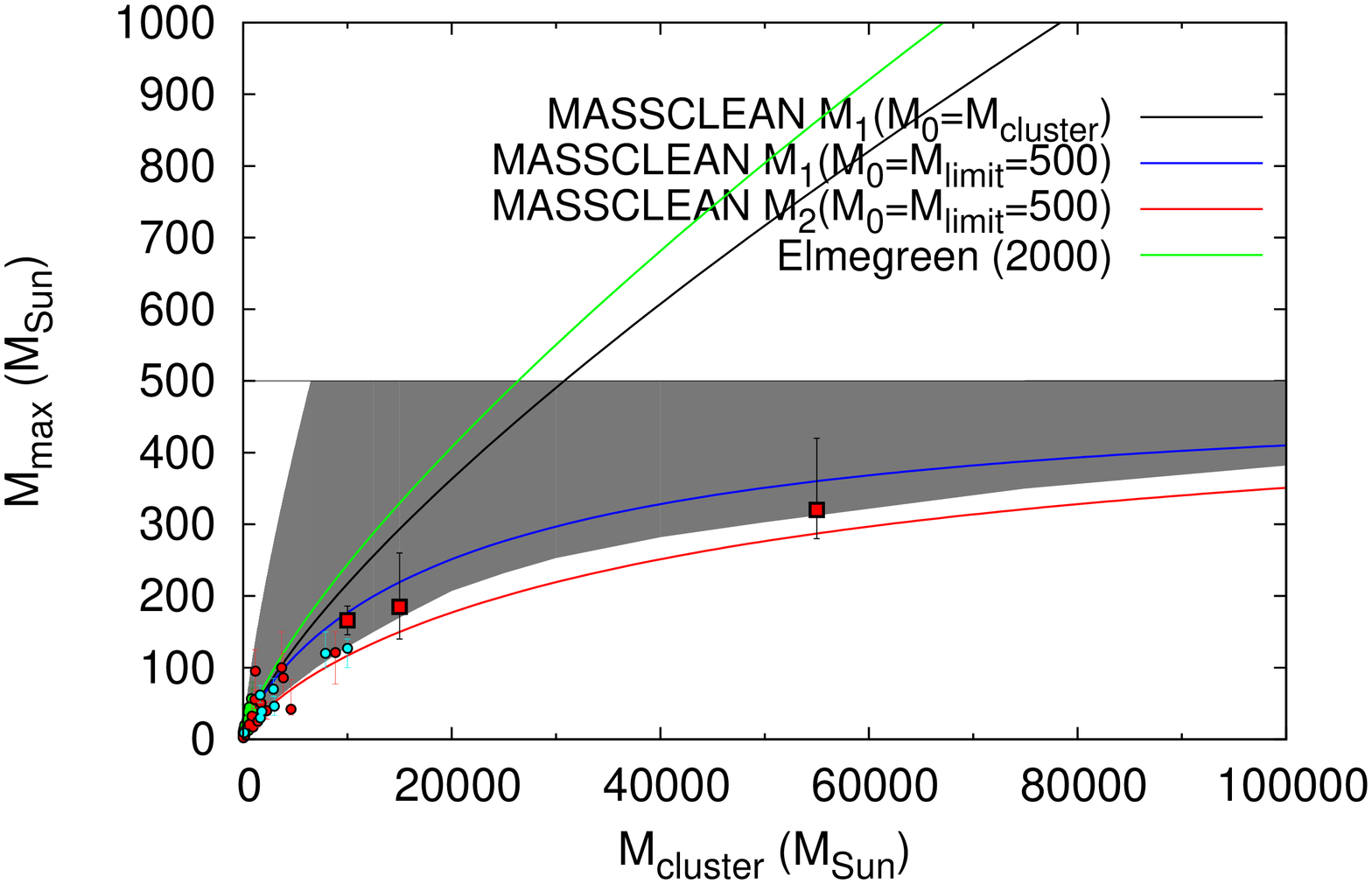}} 
\subfigure[]{\includegraphics[angle=270,width=0.48\textwidth]{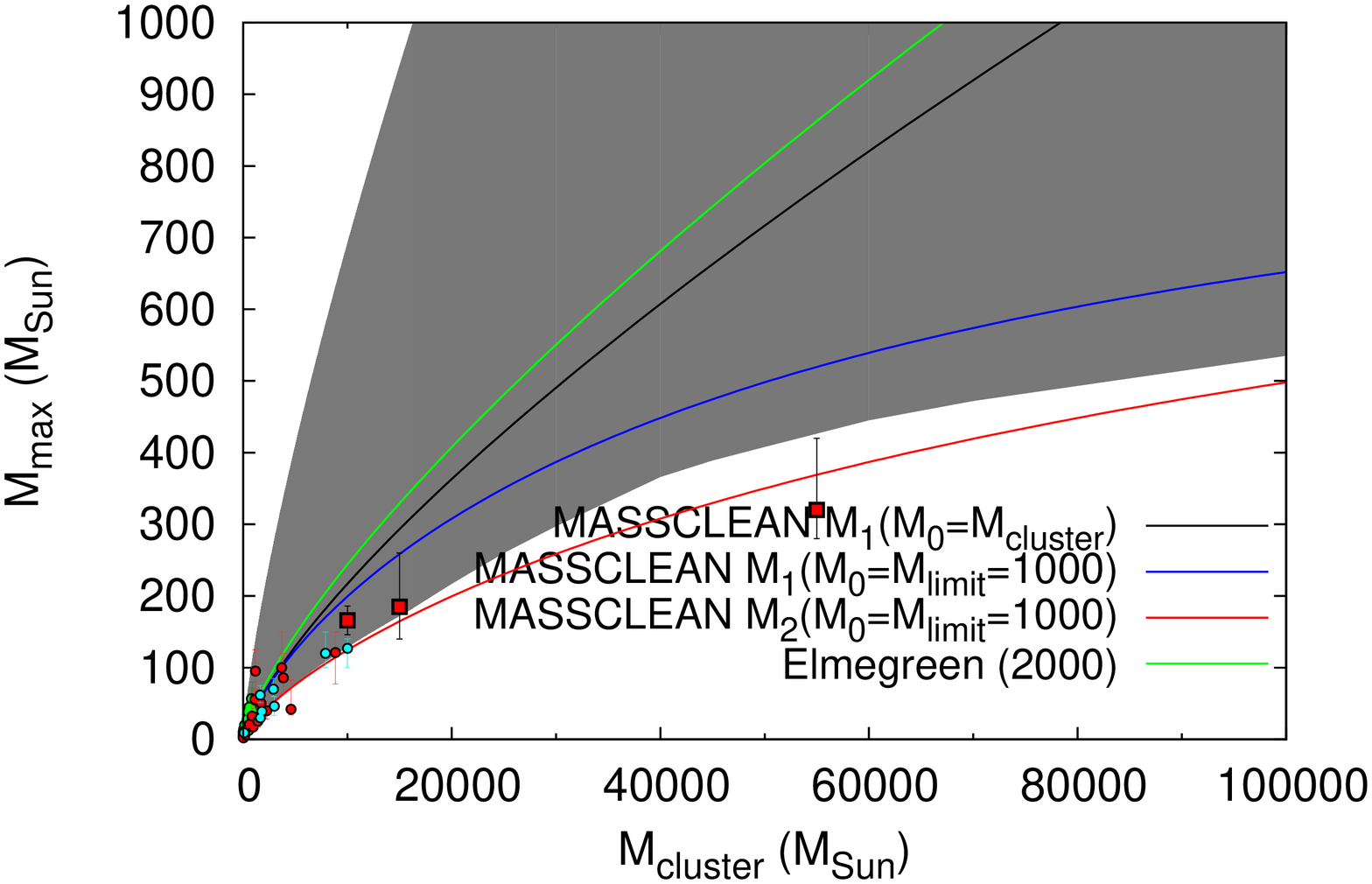}} 
\caption[]{\small The mass of the most massive star, $M_{max}$, versus the total mass of a stellar cluster.  Several analytical limits are given as solid lines.  The gray shaded regions represent MASSCLEAN simulations predicting the mass range of the most massive star expected as a function of stellar cluster mass with upper-limit cutoffs of (a) $150$ $M_{\Sun}$, (b) $300$ $M_{\Sun}$, (c) $500$ $M_{\Sun}$, and (d) $1,000$ $M_{\Sun}$.  The observed maximum stellar mass for three clusters (in mass order: {\it NGC 3603}, {\it Arches} and {\it R136}) as derived by \citeauthor*{crowther2010} \citeyearpar{crowther2010} are shown as red squares. The clusters from \citeauthor*{weidner2010} \citeyearpar{weidner2010} and \citeauthor*{weidner2013} \citeyearpar{weidner2013} are shown as green and red dots, respectively. Other clusters from literature (listed in Table \ref{table1}) are presented as cyan dots.  \normalsize}\label{fig:paper5-04}
\end{figure*}

\begin{figure*}[htp]
\centering
\subfigure[]{\includegraphics[angle=270,width=0.48\textwidth]{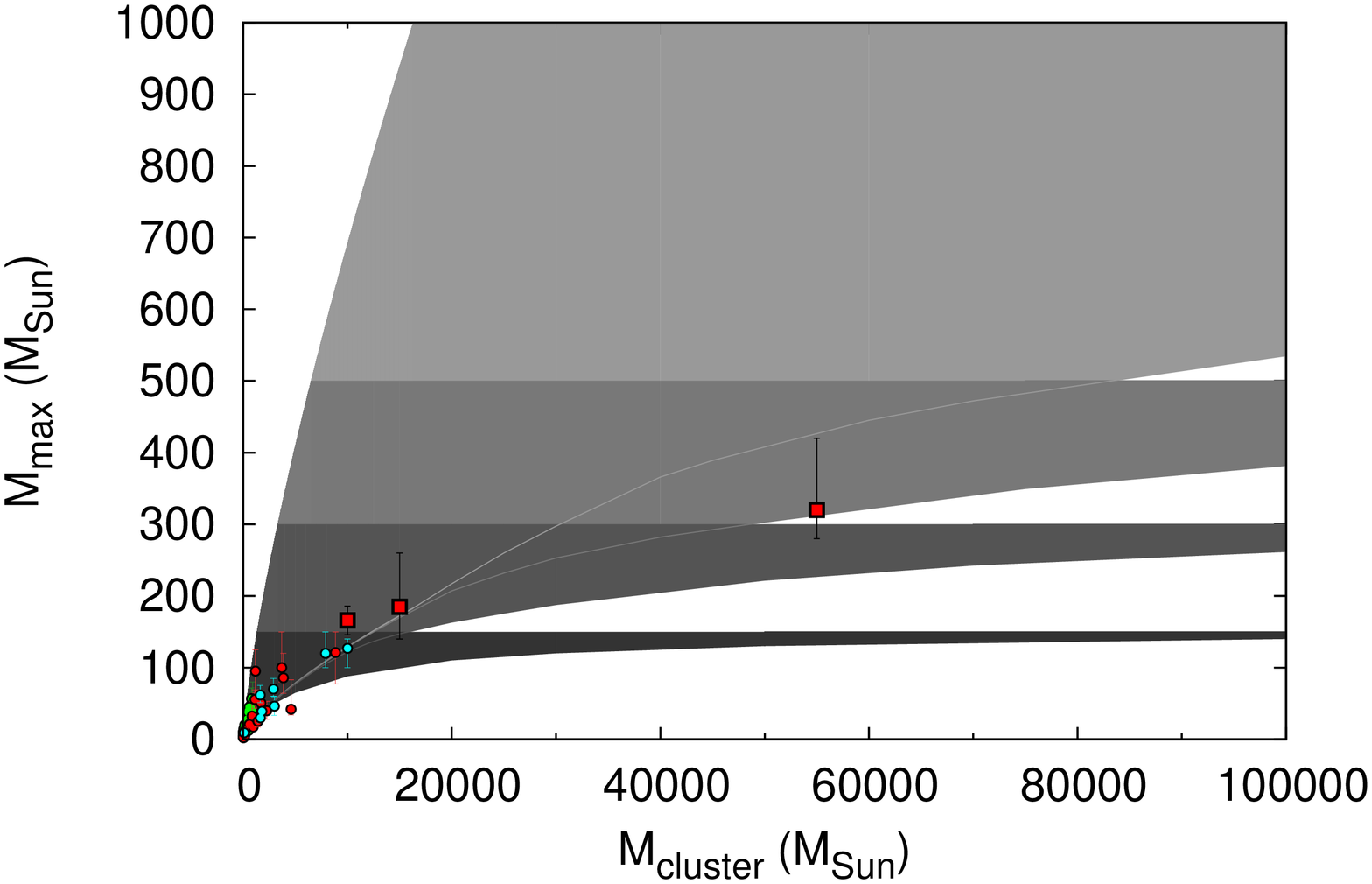}} 
\subfigure[]{\includegraphics[angle=270,width=0.48\textwidth]{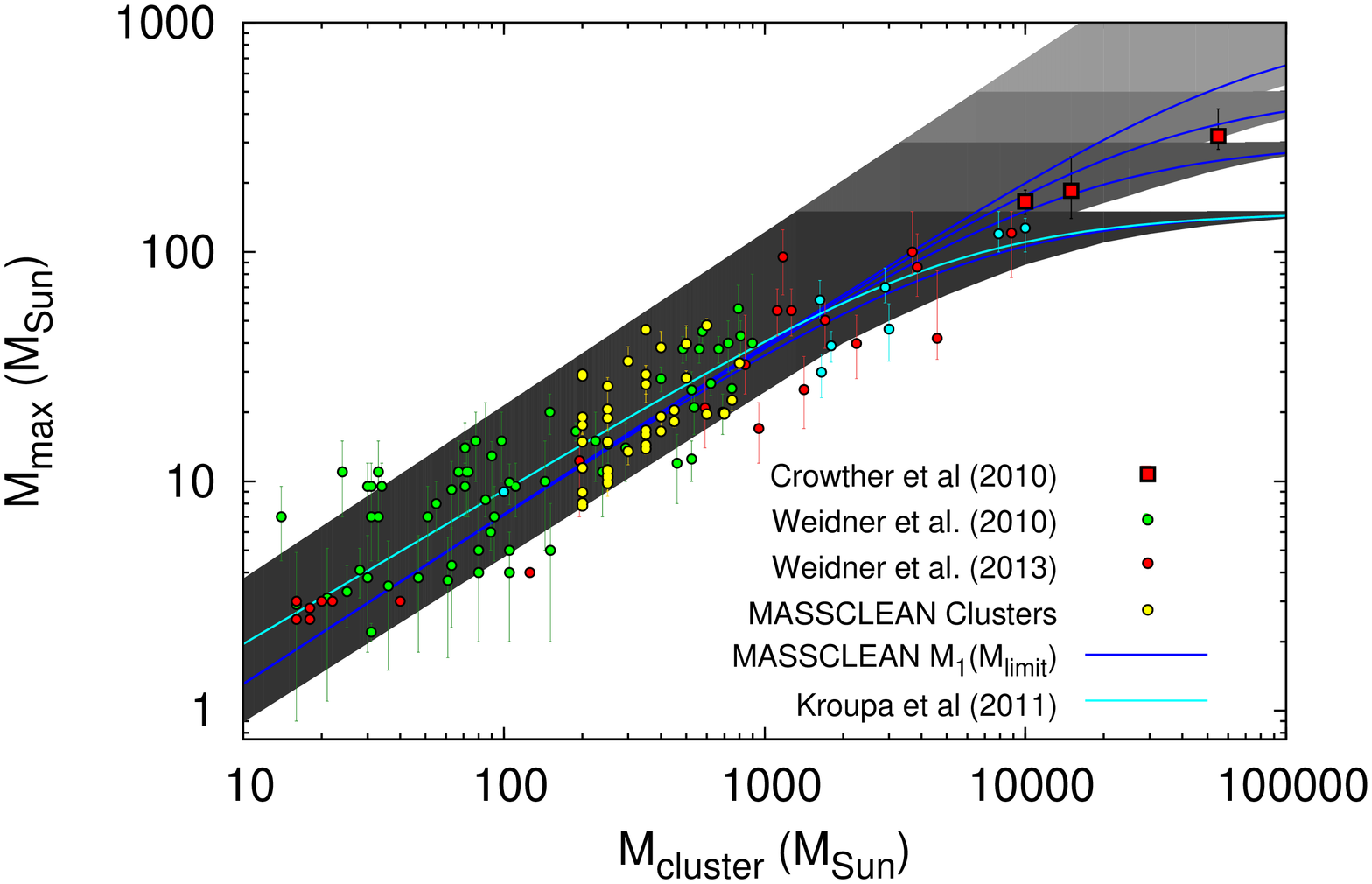}} 
\caption[]{\small Similar to Figure 4, but all four upper-mass limits are given on a single figure with progressive gray tones.  (b) Same as (a), but in log scale. \citeauthor*{crowther2010} \citeyearpar{crowther2010} clusters are shown as red squares. The clusters from \citeauthor*{weidner2010} \citeyearpar{weidner2010} and \citeauthor*{weidner2013} \citeyearpar{weidner2013} are shown as green and red dots, respectively. Other clusters from literature are presented as cyan dots.
\normalsize}\label{fig:paper5-05}
\end{figure*}

\begin{figure}[htp]
\centering
\subfigure[]{\includegraphics[angle=270,width=0.48\textwidth]{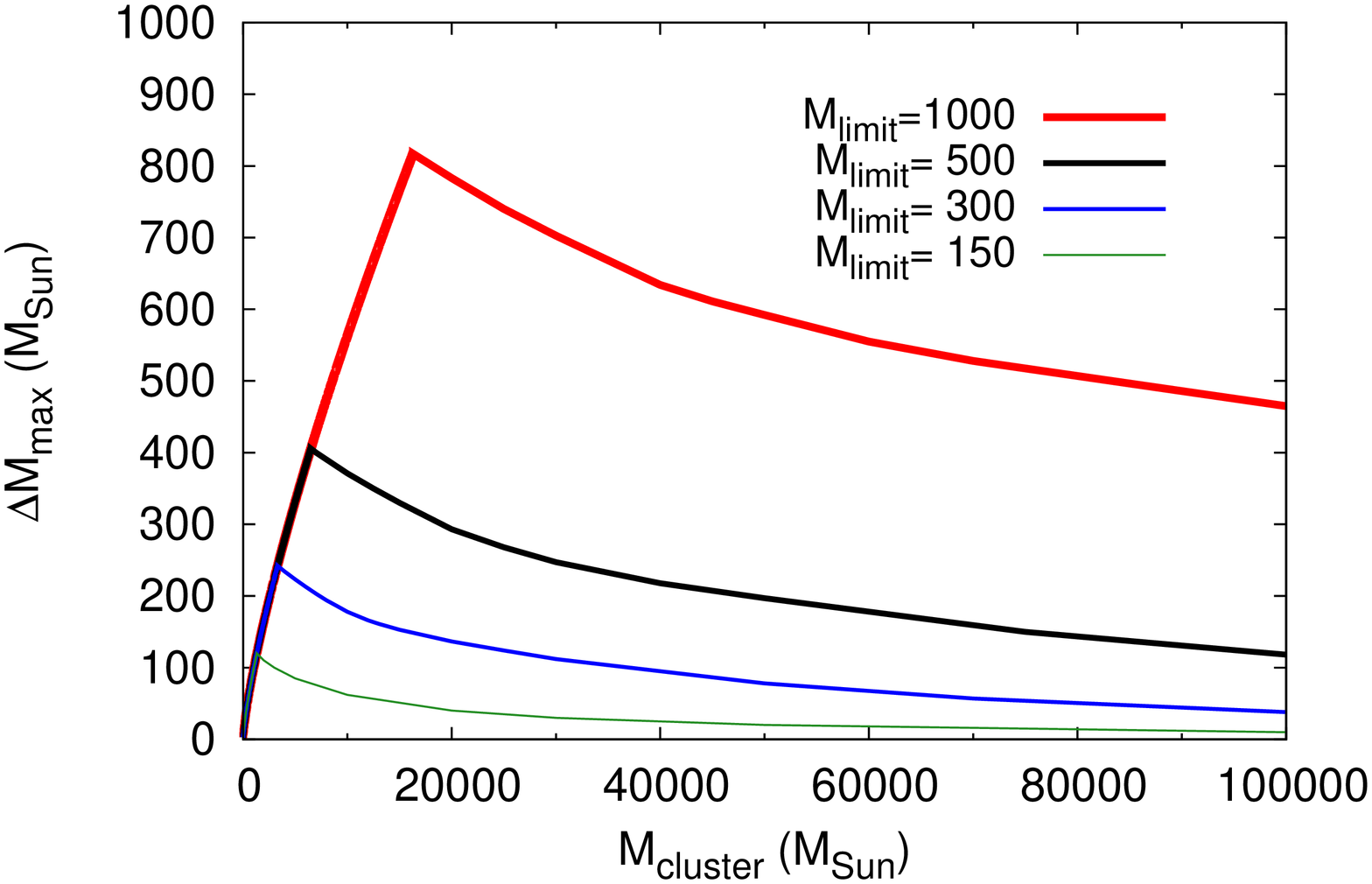}} 
\subfigure[]{\includegraphics[angle=270,width=0.48\textwidth]{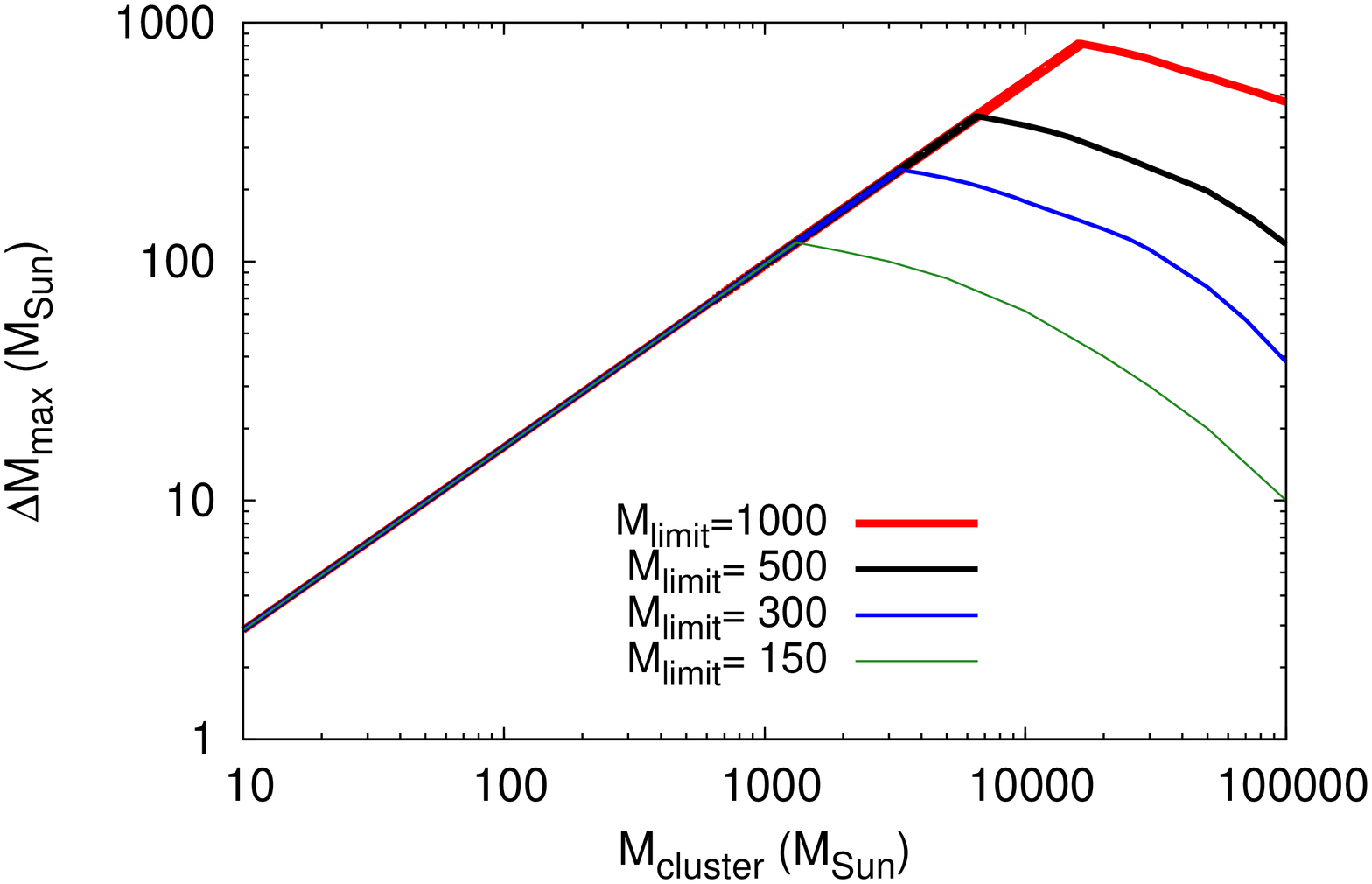}} 
\subfigure[]{\includegraphics[angle=270,width=0.48\textwidth]{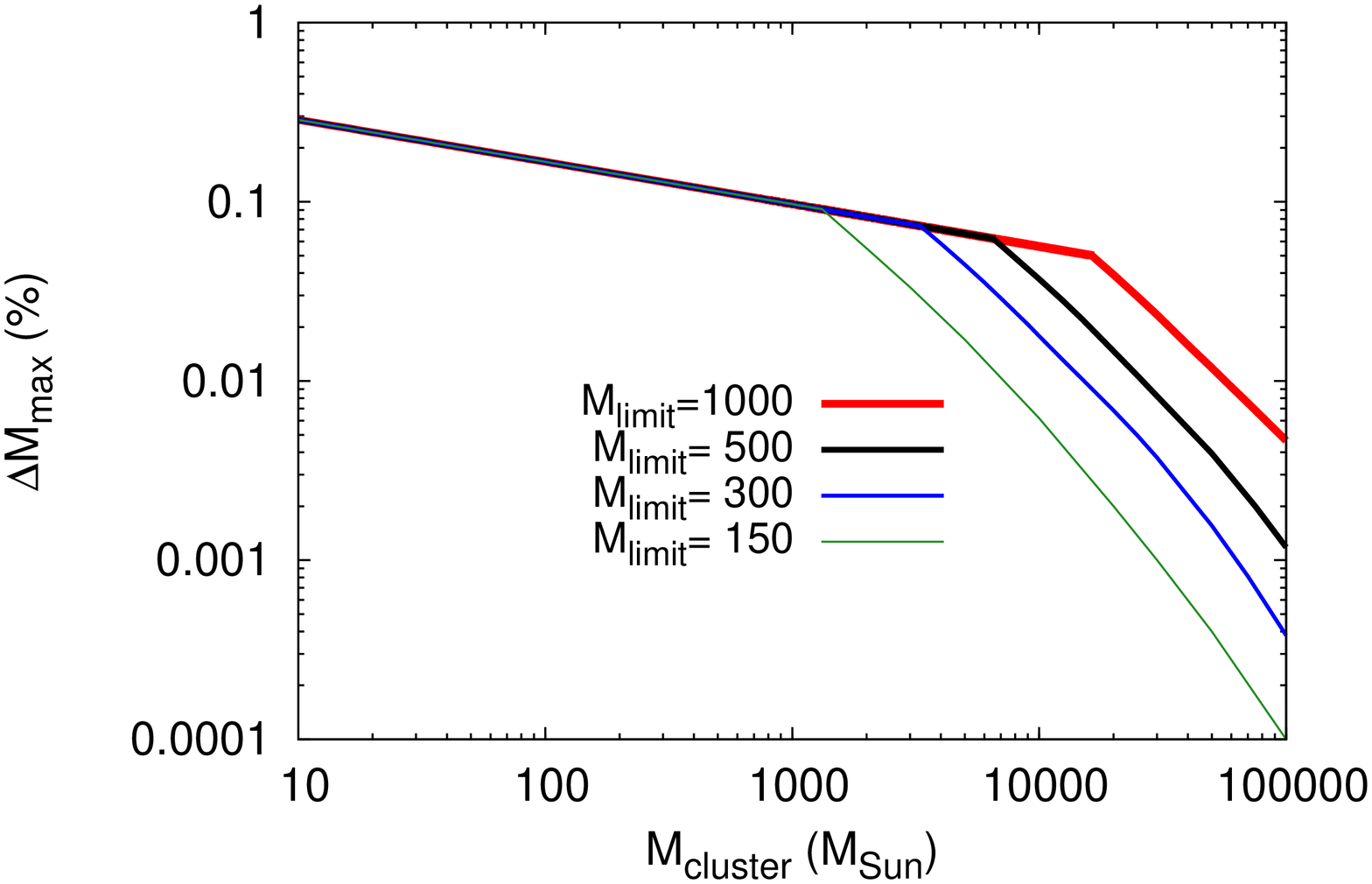}} 
\caption[]{\small  $\Delta M_{max} = M_{max,up} - M_{max,lo}$ versus the total mass of the cluster $M_{cluster}$.  For unsaturated clusters (up to $1,000$ $M_{\Sun}$ clusters) $\Delta M_{max}$ is independent of a stellar upper-mass limit. (a) Linear scale. (b) Logarithmic scale. (c) $\Delta M_{max}$ is presented as a percentage of $M_{cluster}$.  \normalsize}\label{fig:paper5-06}
\end{figure}

Our results from $10$ $million$ Monte Carlo simulations of stellar clusters are presented in Figures \ref{fig:paper5-04} -- \ref{fig:paper5-06}. The mass range of the most massive star, $M_{max}$, as a function of the cluster mass, $M_{cluster}$, is presented in Figure \ref{fig:paper5-04} (a) as the gray-shaded area.  The figure shows the stellar mass range when adhering to the canonical upper mass limit of $M_{limit}=150$ $M_{\Sun}$.  As described before, the \citeauthor*{elmegreen2000} \citeyearpar{elmegreen2000} maximum mass (green line, Equation (\ref{eq:18})) and our Equation (\ref{eq:21}) (black line) will diverge at large enough cluster mass.  In blue we plot Equation (\ref{eq:23}), with $M_{1}(M_{limit}=150$ $M_{\Sun})$.  This resembles very well the canonical form of \citeauthor*{kroupa2011} \citeyearpar{kroupa2011}, displayed as the cyan line. Note the blue line is not aligned with the bottom of the gray-shaded $M_{max}$ region. This is because, as we have already described, sometimes the most massive star does not fall inside the massive star bin.  We show $M_{2}(M_{limit})$ as the red line.   This represents the rare, but very real situation where the two most massive stars share the two top mass bins (Equation (\ref{eq:26})).  This red line lies just below the lower limit of the $M_{max}$ gray-shaded range. 

In Figure \ref{fig:paper5-04} (a) we present as the magenta line the latest fit of data from literature presented by \citeauthor*{weidner2013} \citeyearpar{weidner2013} assuming the $150$ $M_{\Sun}$ {\it canonical limit} (their Equation (1)). This line is significantly lower than all of the other variation presented in \ref{fig:paper5-04} (a) due to the overestimation of $M_{cluster}$ at the high mass end. For example, the {\it Arches} cluster is listed with over $77,000$ $M_{\Sun}$ mass in \citeauthor*{weidner2013} \citeyearpar{weidner2013}, while the extensive study from \citeauthor*{clarkson2012} \citeyearpar{clarkson2012} determine only a $15,000$ $M_{\Sun}$. (This is the value also used in our plots, see below). Similarly \citeauthor*{crowther2010} \citeyearpar{crowther2010} presents a $55,000$ $M_{\Sun}$ for {\it R136}, which is significantly lower than over $200,000$ $M_{\Sun}$ in \citeauthor*{weidner2013} \citeyearpar{weidner2013}.

Figures \ref{fig:paper5-04} (b), (c), and (d) are similar to Figure \ref{fig:paper5-04} (a), but for $M_{limit}=300, 500,$ and $1,000$ $M_{\Sun}$, respectively. The {\it canonical} $M_{max}$$-$$M_{cluster}$ relation of \citeauthor*{kroupa2011} \citeyearpar{kroupa2011} is only available for $M_{limit}=150$ $M_{\Sun}$, but its variation is described well by the blue line (our Equation (\ref{eq:23})) in these three figures, with the respective $M_{limit}$.  

What should be immediately obvious from Figures \ref{fig:paper5-04} (a)--(d) is that despite millions of simulations, there is a hard upper limit on the most massive star expected to be formed, and it is a function of cluster mass.  Even if you sampled a million $500$ $M_{\Sun}$ clusters, our simulations indicate a $100$ $M_{\Sun}$ star will never be formed.  Yet, $100$ $M_{\Sun}$ stars are predicted to form in clusters with just $1,000$ $M_{\Sun}$, though these clusters can not form $150$ $M_{\Sun}$ stars, and so forth.  What is coming out naturally from our simulations is that the lack of very high mass stars in low mass clusters is not a size of sample effect. The upper limit on the mass of the most massive star seen being tied to the mass of the initial cluster is entirely predicted from our simulations when we properly populate the cluster's IMF.

On Figures \ref{fig:paper5-04} (a)--(d), as a critical observational reference, we also present the location of the most massive stars for the three massive clusters as given by \citeauthor*{crowther2010} (\citeyear{crowther2010}).  The three stars are shown as red squares representing: $166 \pm 20$ $M_{\Sun}$ ({\it NGC 3603}), $185^{+75}_{-45}$ $M_{\Sun}$ ({\it Arches}), and $320 \pm 40$ $M_{\Sun}$ ({\it R136}). We also show the clusters with the mass less than $1,000$ $M_{\Sun}$ from \citeauthor*{weidner2010} \citeyearpar{weidner2010} as green dots. Additional low mass clusters and young clusters (1 Myr) above $1,000$ $M_{\Sun}$ from \citeauthor*{weidner2013} \citeyearpar{weidner2013} are presented as red dots. Additional clusters from literature (\citeauthor*{chene2012} \citeyear{chene2012}; \citeauthor*{davies2012} \citeyear{davies2012}; \citeauthor*{martins2010} \citeyear{martins2010}; \citeauthor*{ascenso2007} \citeyear{ascenso2007}; \citeauthor*{hur2012} \citeyear{hur2012}; \citeauthor*{crowther2012} \citeyear{crowther2012}; \citeauthor*{bonatto2006} \citeyear{bonatto2006}; \citeauthor*{deharveng2009} \citeyear{deharveng2009}) are shown as cyan dots. All of these clusters are listed in the Table \ref{table1}.

In Figure \ref{fig:paper5-05} (a) we present all of the gray-scale ranges from Figures \ref{fig:paper5-04} (a)--(d) on the same plot, using four different shades of gray. Clusters from the literature are presented similarly to Figure \ref{fig:paper5-04}. The Figure \ref{fig:paper5-05} shows that the \citeauthor*{crowther2010} \citeyearpar{crowther2010} clusters are consistent, within the error bars, with an upper stellar mass limit, $M_{limit}$, in the $300-1000$ $M_{\Sun}$ range.  But \citeauthor*{kroupa2011} \citeyearpar{kroupa2011}, citing results from \citeauthor*{banerjee2012} (\citeyear{banerjee2012}; \citeyear{banerjee2012b}), claim these very massive stars could be so-called {\it supercanonical stars}, formed by the collision of massive binary stars. They use this to claim a $M_{limit}$ in the range of $150$ $M_{\Sun}$ can not be excluded despite the presence of these super-massive stars.  Regretfully, this is a classic circular argument.  Moreover, there is no observational evidence that stars more massive than $150$ $M_{\Sun}$ must form differently, such as from mergers.

A very interesting feature of the dispersion in $M_{max}$ determined using the MASSCLEAN simulations should be noted in Figures \ref{fig:paper5-05} (a) and (b). The range of $M_{max}$ is independent of $M_{limit}$ as long as the mass of the cluster is smaller than the value where the upper stellar mass limit could be reached, {\sl i.e.}, what \citeauthor*{kroupa2011} \citeyearpar{kroupa2011} defines to be {\it unsaturated clusters}). This is a result that would be expected, and we consider it a validation of the MIMFS method. For example, from our {\it 10 million} Monte Carlo simulations, the mass of the most massive star of a $1,000$ $M_{\Sun}$ cluster is less than $120$ $M_{\Sun}$.  There is no reason the dispersion in $M_{max}$ for clusters with masses lower than $1,000$ $M_{\Sun}$ should depend in any way on a $M_{limit}$, if the limit is higher than $120$ $M_{\Sun}$. 

Everything from Figure \ref{fig:paper5-05} (a) is displayed on Figure \ref{fig:paper5-05} (b), but using a logarithmic scale and including many more observational points representing real clusters. Here it is borne out:  the most massive star seen never exceeds the range of values predicted from our simulations over a cluster mass range of nearly two orders of magnitude.   Such a relationship can only be proven to exist by studying these low mass, unsaturated clusters, where the maximum stellar mass is not being limited by any additional, outside stellar physics limit (indicated by a $M_{lim}$ value). 
 Once again, it is obvious that the dispersion in $M_{max}$ is independent of $M_{limit}$ for {\it unsaturated clusters} and that the upper stellar mass limit should be at least $150$ $M_{\Sun}$. Further, Figure \ref{fig:paper5-05} (b) confirms that clusters with $M_{cluster}<1,000$ $M_{\Sun}$ are not affected by the choice for $M_{limit}$ in the $150-1,000$ $M_{\Sun}$ range. For all these reasons, low mass clusters are the ideal candidates to study $M_{max}$. 
 This is why we included in the figure the observed values from the low-mass clusters from \citeauthor*{weidner2010} \citeyearpar{weidner2010}, presented as green dots. Additional clusters from \citeauthor*{weidner2013} \citeyearpar{weidner2013} are presented as red dots, and other clusters from literature are shown as cyan dots. All of these clusters are consistent with the dispersion range of $M_{max}$ we determined. 

The \citeauthor*{crowther2010} \citeyearpar{crowther2010} clusters are presented as red squares in Figure \ref{fig:paper5-05}, but they are outside the low mass range, which is our main focus. The {\it canonical} $M_{max}$$-$$M_{cluster}$ relation from our $M_{1}(M_{limit})$ and \citeauthor*{kroupa2011} \citeyearpar{kroupa2011} are presented as the blue lines (with four different values for $M_{limit}$) and the single cyan line, respectively. We also include 40 MASSCLEAN clusters, as yellow dots. They will be described in the next section.

For yet another view of the MIMFS simulation results, we present the dispersion range, the difference between the upper and the lower limits of $M_{max}$, $\Delta M_{max} = M_{max,up}- M_{max,lo}$, in Figure \ref{fig:paper5-06} (a), for all four values of the $M_{limit}$ discussed above versus the cluster total mass. Since the range of $M_{max}$ is independent of $M_{limit}$ for {\it unsaturated clusters}, $\Delta M_{max}$ is also independent. The Figure \ref{fig:paper5-06} (b) is the same as Figure \ref{fig:paper5-06} (a), only presented in the logarithmic scale. The Figure \ref{fig:paper5-06} (c) is again the same values plotted, only now we are showing the  $\Delta M_{max}$ as a percentage of $M_{cluster}$, in a logarithmic scale.

\section{The Most Massive Star in Stellar Clusters Derived from Integrated Magnitudes and Colors}

The degree to which the integrated colors and magnitudes of stellar clusters can be expected to vary away from the mean SSP-model prediction is strongly anti-correlated to the cluster's mass (e.g. \citeauthor*{paper1} \citeyear{paper2}, \citeyear{paper3}; \citeauthor*{paper4} \citeyear{paper4}).  In other words, the most massive star in a cluster's mass distribution has an increasingly prominent influence in the magnitude and colors of lower-mass clusters.   It is our goal to exploit this observed dispersion to estimate the mass of the most massive star in a sample of low-mass stellar clusters of known age and mass, using nothing but the integrated broad-band magnitudes of the clusters.

To do this, we investigated the variation of $U,B,V$ colors as a function of the mass of the most massive star in the distribution, $M_{max}$. We performed $25$ $million$ MASSCLEAN Monte Carlo simulations for clusters in the $200-1,000$ $M_{\Sun}$ range. We used $M_{cluster}  = 1,000 M_{\Sun}$ as the upper limit because in this range the dispersion in the maximum stellar mass, $M_{max}$, is independent of the stellar upper mass limit, $M_{limit}$.  As described in \S 4, these clusters are expected to be {\it unsaturated}.  We used the Kroupa-Salpeter IMF (Equation (\ref{eq:3})), and the Padova stellar evolutionary models (\citeauthor*{padova2008} \citeyear{padova2008}; \citeauthor*{padova2010} \citeyear{padova2010}), for $Z=0.008$ metallicity.   In this way, we have created a special version of the MASSCLEAN{\fontfamily{ptm}\selectfont \textit{colors}} database (\citeauthor*{paper3} \citeyear{paper3}; \citeauthor*{paper4} \citeyear{paper4}) which contains mass, age, $U,B,and V$, but {\sl also now} $M_{max}$\footnote{This is the {\it initial} mass of the most massive star in the distribution, unaffected by the stellar evolution.} for each cluster.

\begin{figure*}[htp]
\centering
\subfigure[]{\includegraphics[angle=270,width=0.48\textwidth]{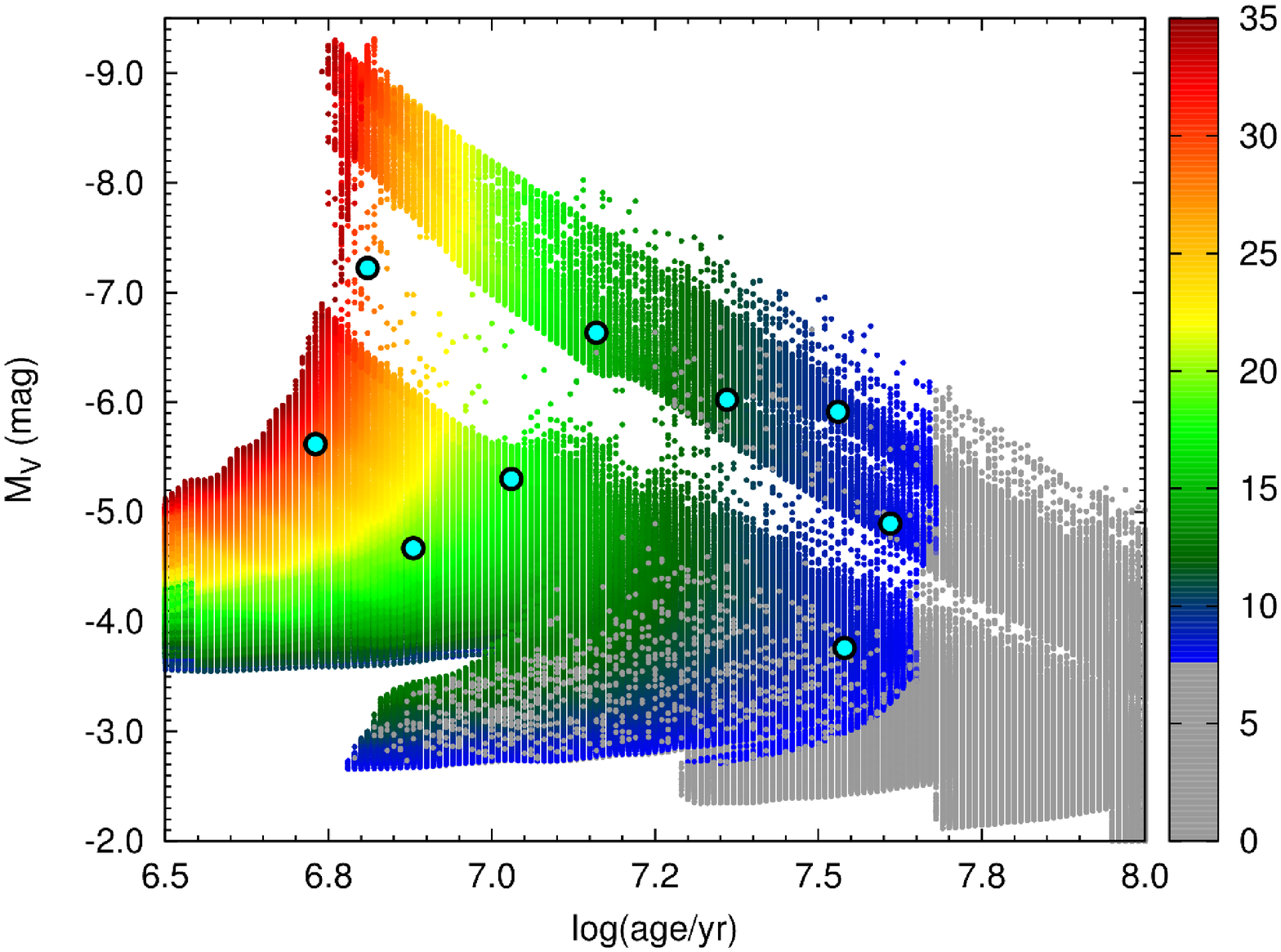}} 
\subfigure[]{\includegraphics[angle=270,width=0.48\textwidth]{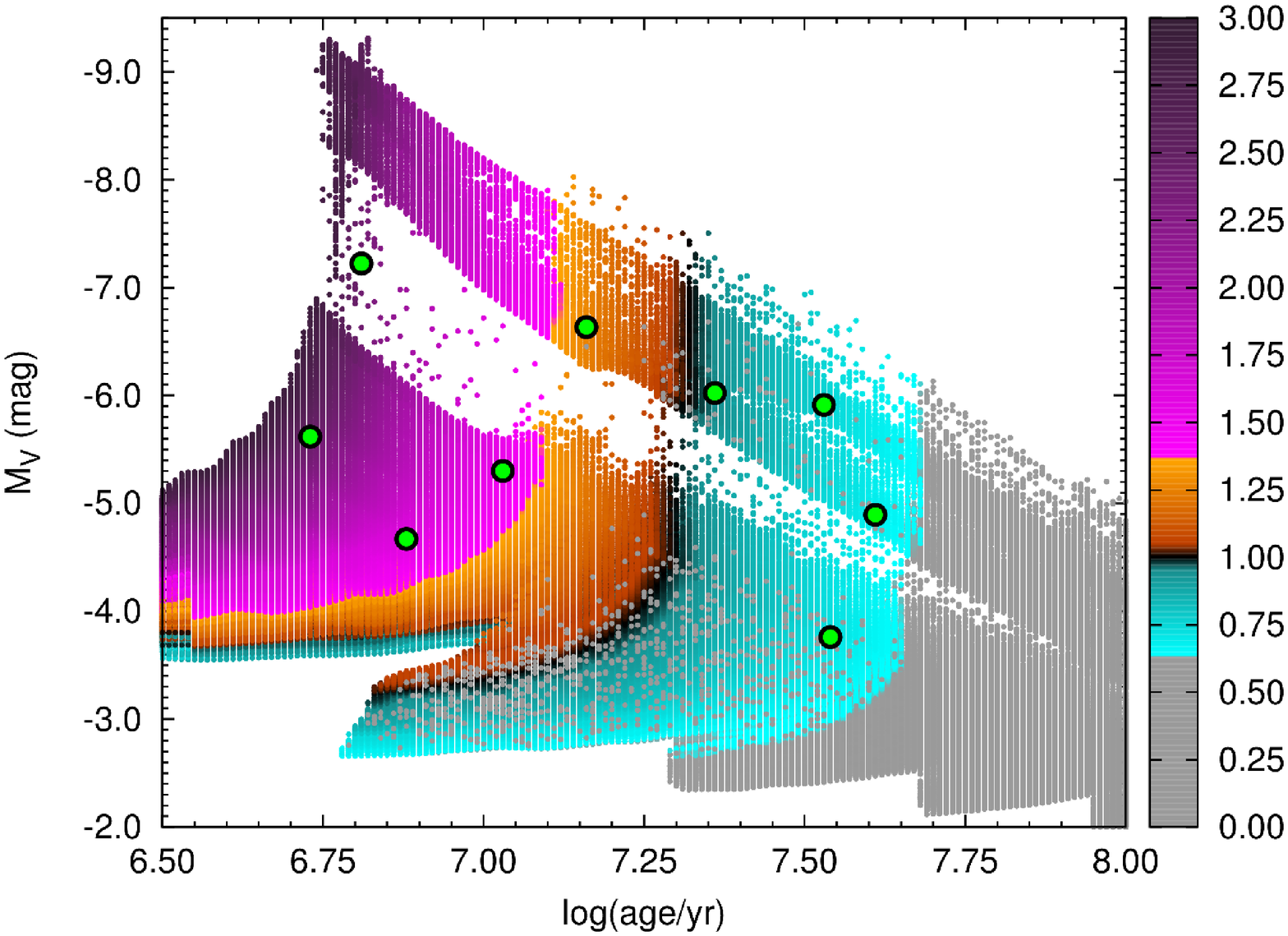}} 
\subfigure[]{\includegraphics[angle=270,width=0.48\textwidth]{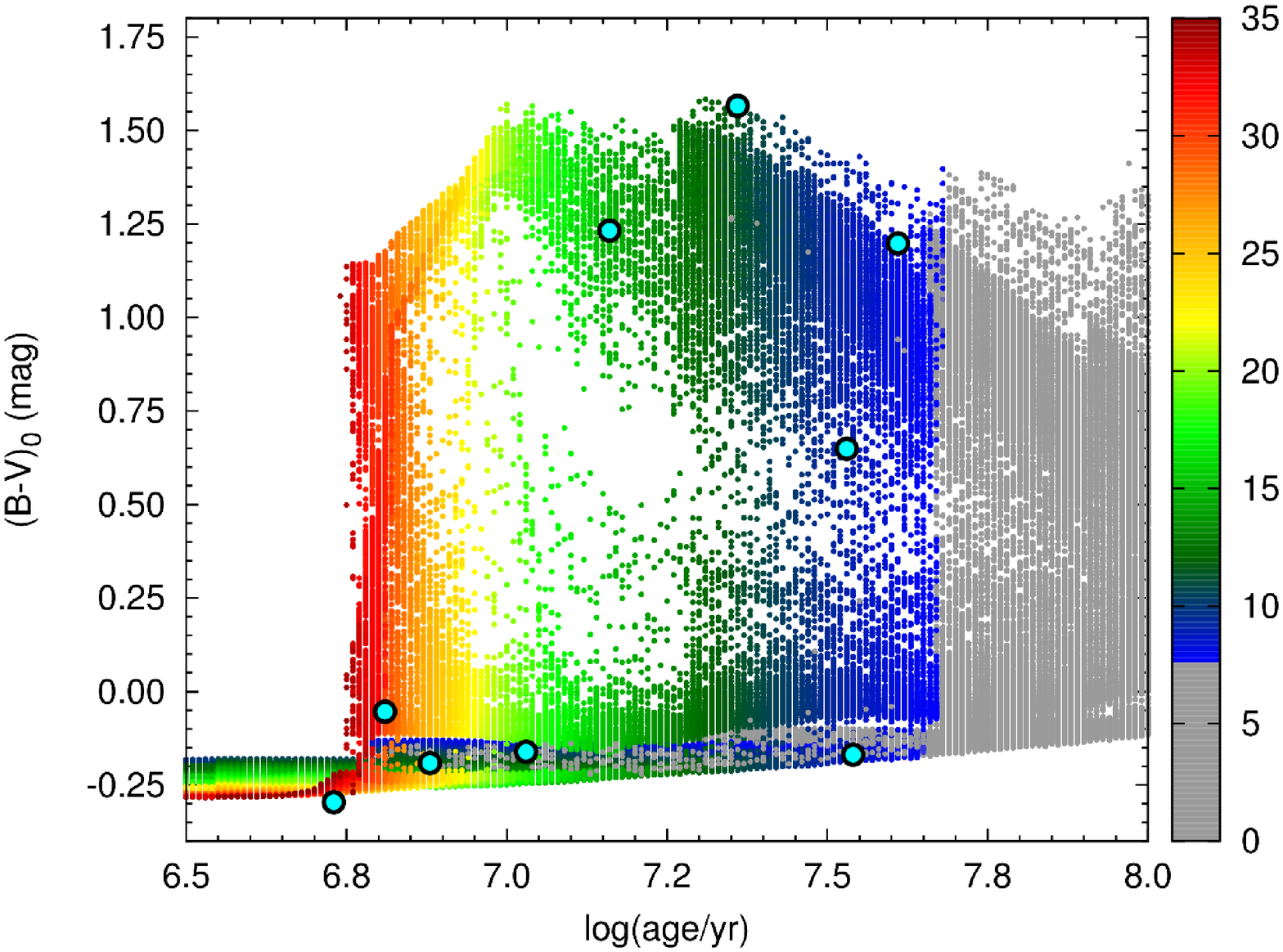}} 
\subfigure[]{\includegraphics[angle=270,width=0.48\textwidth]{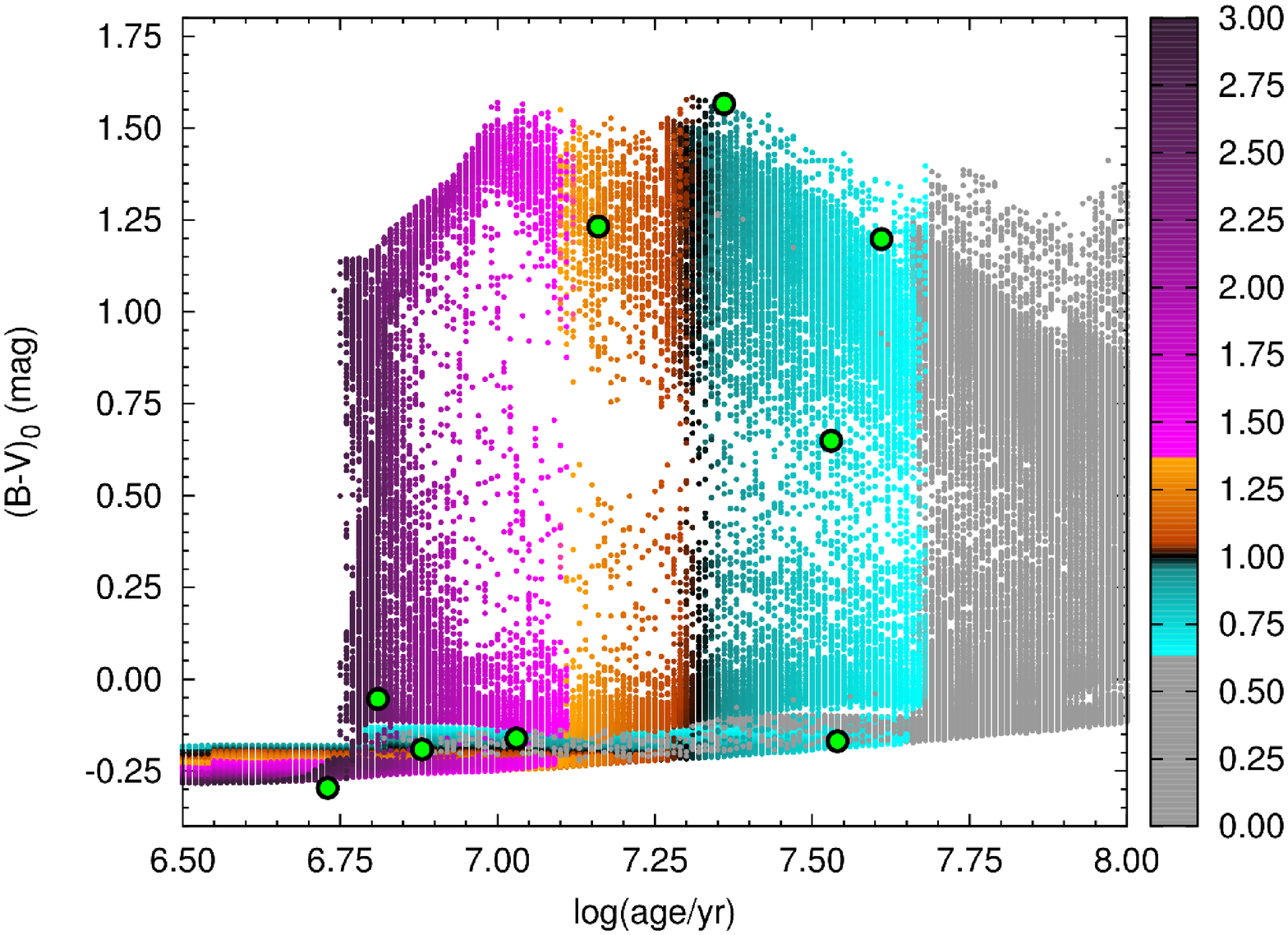}} 
\subfigure[]{\includegraphics[angle=270,width=0.48\textwidth]{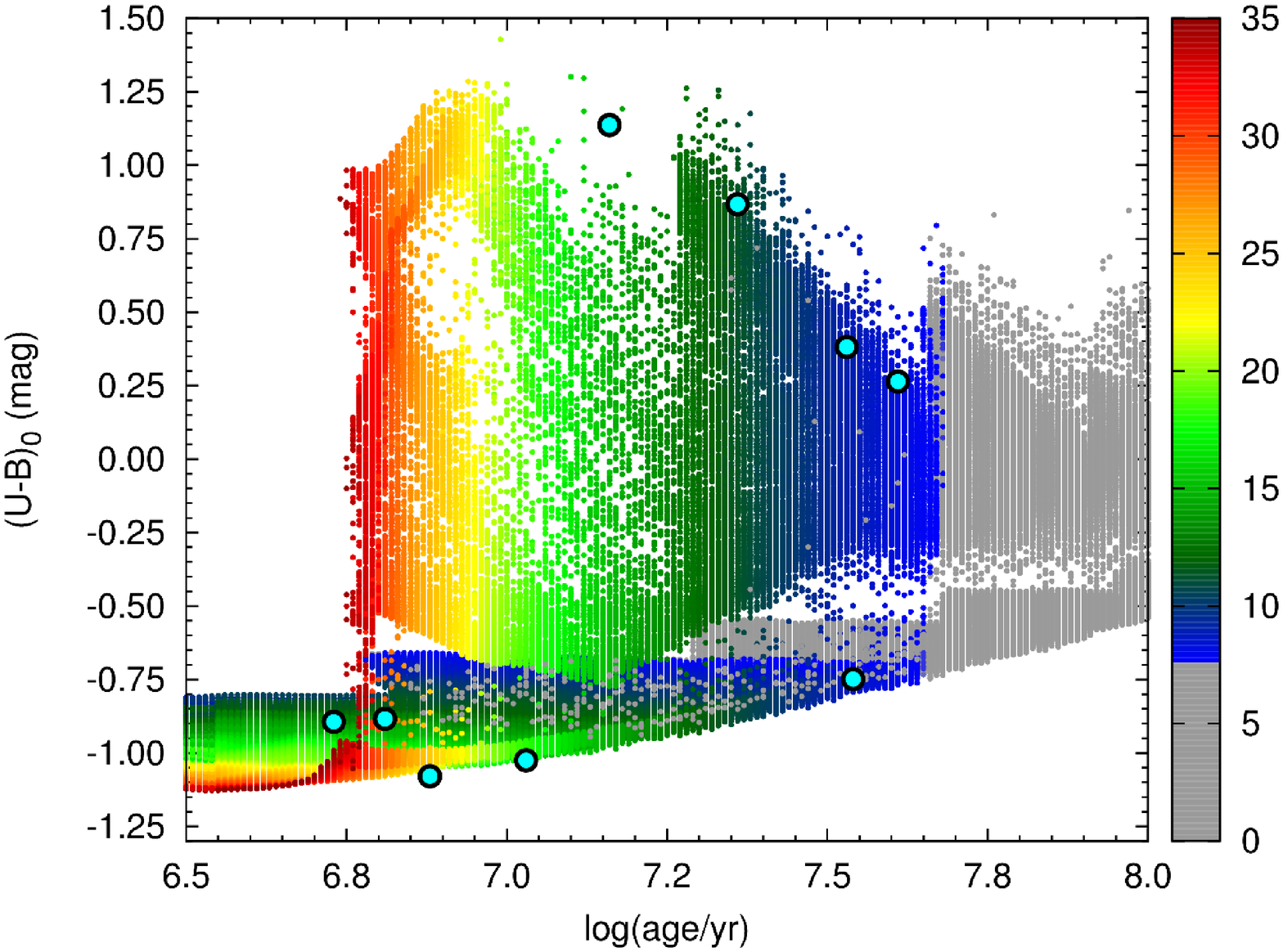}} 
\subfigure[]{\includegraphics[angle=270,width=0.48\textwidth]{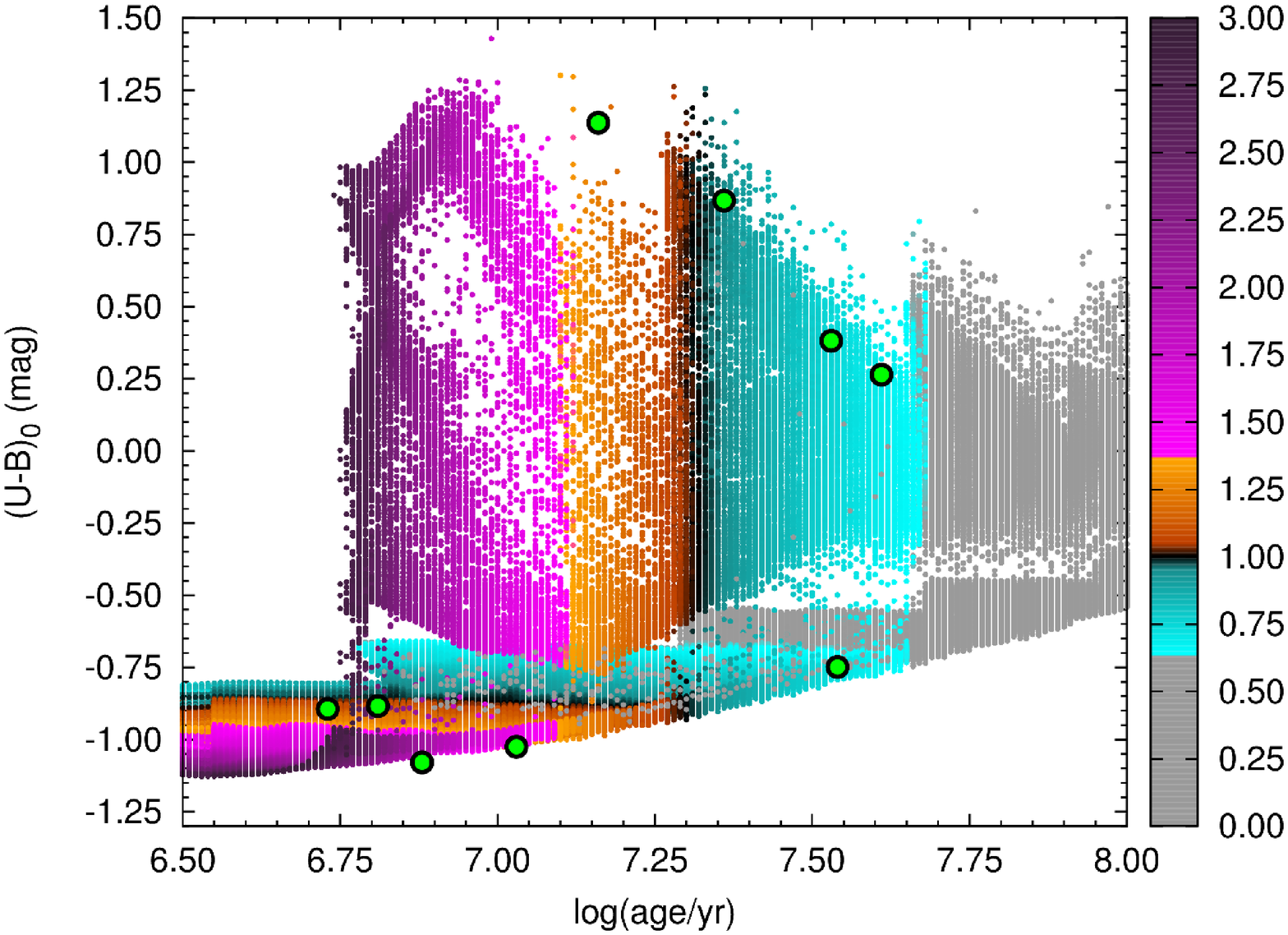}} 
\caption[]{\small A subset of our $25$ $million$ Monte Carlo simulations, for $M_{cluster}=200$ $M_{\Sun}$. The cyan dots ({\it Left}), and green dots ({\it Right}) are MASSCLEAN clusters from \citeauthor*{paper4} \citeyear{paper4}.  {\it Left Panels}: Dots are color-coded to show the integrated color or magnitude of a 200 $M_{max}$ cluster with that $M_{max}$, the mass of the most massive star, with time. The clusters displayed in gray already lost their most massive star. {\it Right Panels}: Dots are color-coded to show the ratio $M_{max}/M_{1}$, and the gray clusters already lost the most massive star. \normalsize}\label{fig:paper5-07}
\end{figure*}

In Figure \ref{fig:paper5-07}  we show an example of the variation of $M_{V}$, $(B-V)_{0}$, and $(U-B)_{0}$ for $M_{cluster}=200$ $M_{\Sun}$.  In the three {\it left} panels, the small dots are rainbow-coded to indicate the maximum mass, $M_{max}$. Blue, green, yellow, orange, and red show the value of $M_{max}$ for a cluster with the given absolute magnitude and age.  The gray region of the diagram is filled with small dots representing clusters that have already lost their most massive star in the distribution due to evolution.  The {\it right} three panels are also color-coded, but show the ratio between the mass of the most massive star in the distribution and the {\it canonical} value, $M_{max} / M_{1}(M_{limit})$. The {\it canonical value} is displayed in black (ratio = 1.0) and the $\pm 35 \%$ values around the {\it canonical value} are shown in shades of cyan and orange, respectively. Higher values of this ratio are presented in different shades from magenta to dark-magenta. All the clusters that are too old and have already lost their most massive star are again displayed in gray. 

The influence of the most massive star on the integrated properties of a cluster is most obvious in Figures \ref{fig:paper5-07} (a) and (b). For a given age, the integrated $M_{V}$ magnitude of a low-mass cluster will greatly increase with $M_{max}$. 

Also shown in Fig \ref{fig:paper5-07} are nine clusters with cluster masses, $M_{cluster}=200$ $M_{\Sun}$ from \citeauthor*{paper4} \citeyearpar{paper4}.  These are given as cyan dots in the left panels and as green dots in the right panels. These clusters have well constrained age and mass, determined from our mass-dependent SSP MASSCLEAN models (\citeauthor*{paper3} \citeyear{paper3}; \citeauthor*{paper4} \citeyear{paper4}). They were selected to be young enough to still contain their most massive star (i.e. to be on the {\it colored} part of the plots presented in Figure \ref{fig:paper5-07}).  It can be seen that the clusters position on the color-age planes presented in Figure \ref{fig:paper5-07} is related to the most massive star in the cluster.  One can see how this might be used to estimate the value of the most massive star in a cluster of known mass and age, based on an analysis of its broadband colors. 

\begin{figure*}[htp]
\centering
\subfigure[]{\includegraphics[angle=270,width=0.48\textwidth]{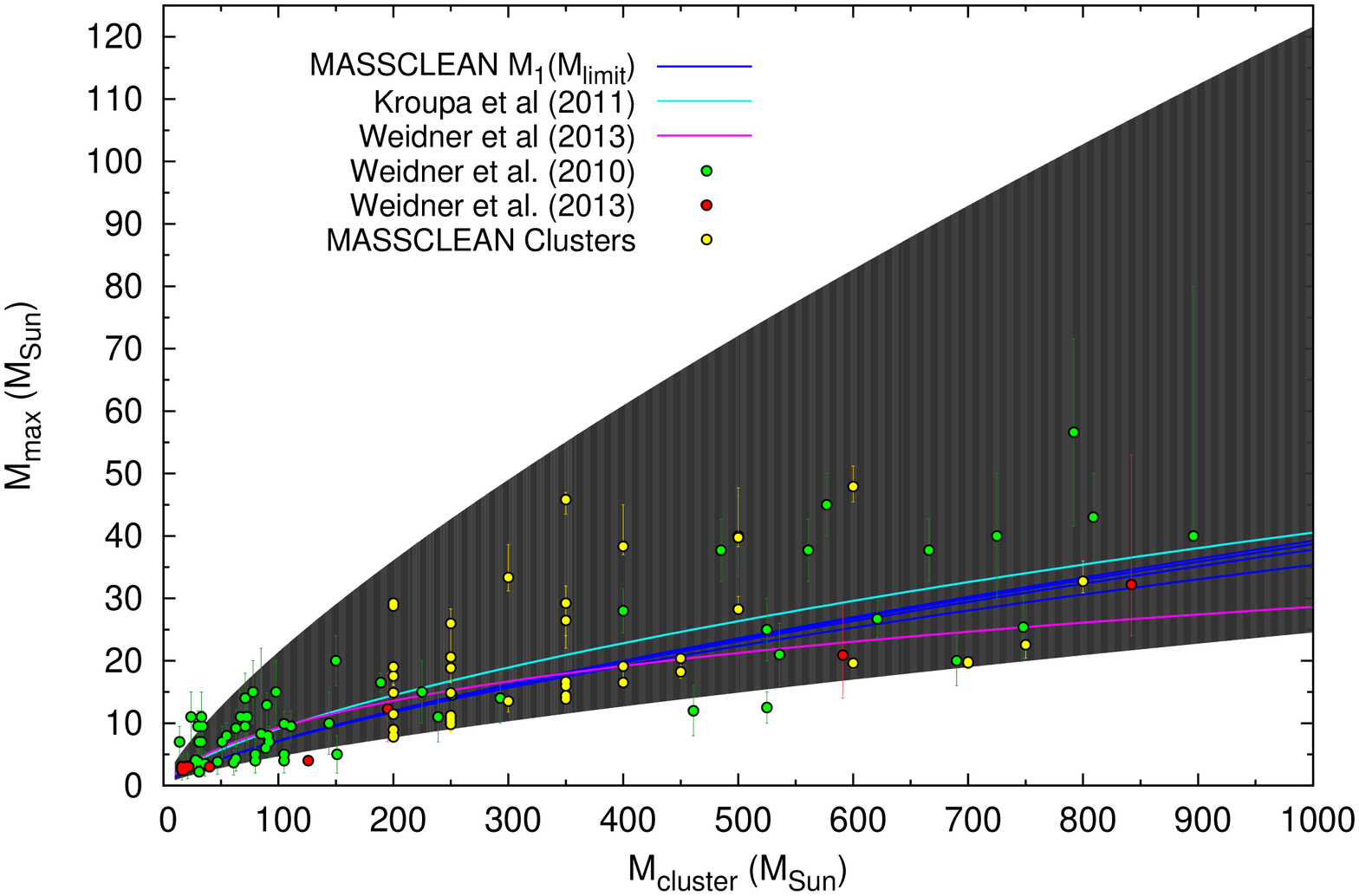}} 
\subfigure[]{\includegraphics[angle=270,width=0.48\textwidth]{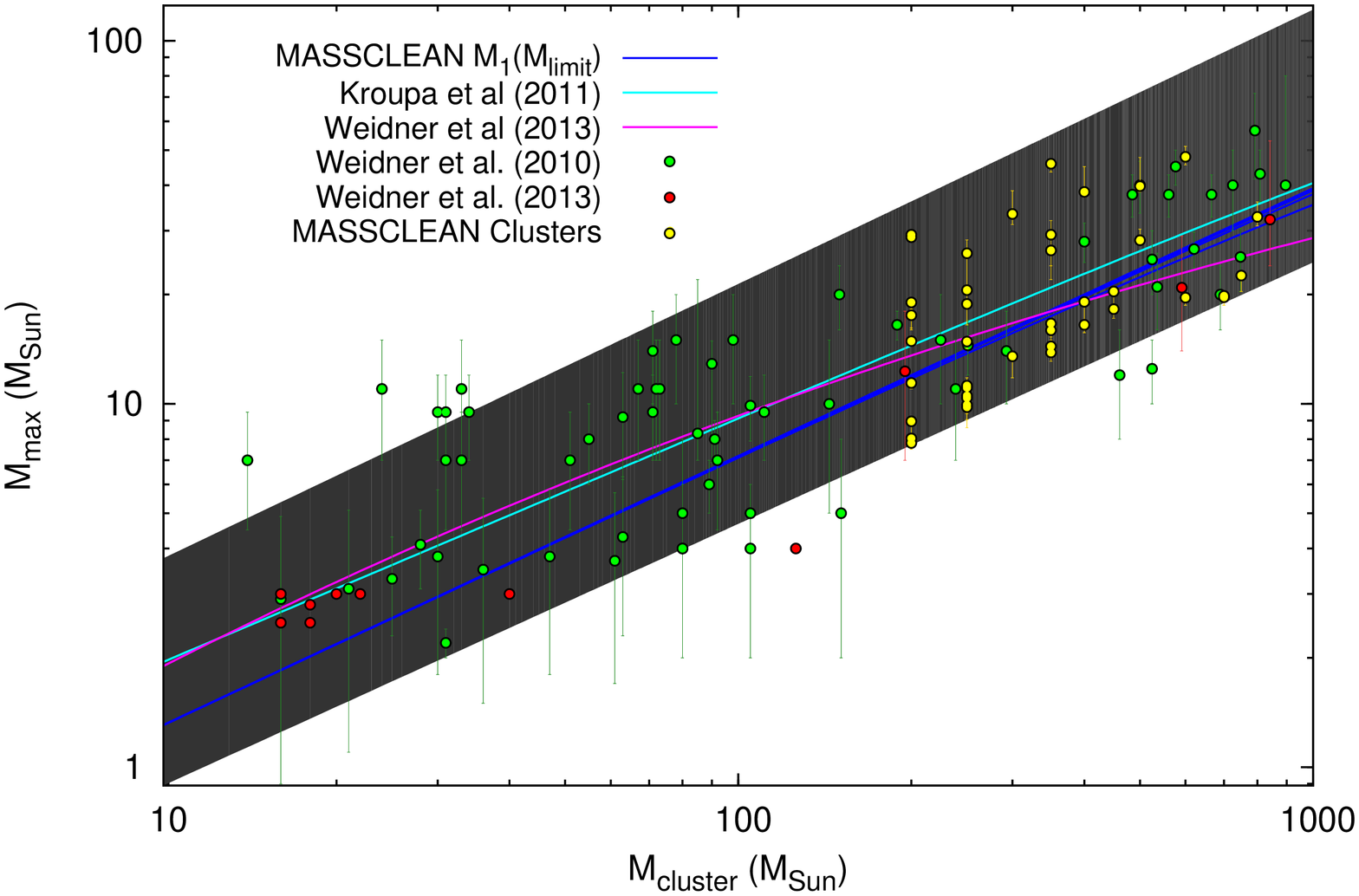}} 
\caption[]{\small Maximum stellar mass, $M_{max}$, versus stellar cluster total mass, $M_{cluster}$, similar to Figure 5 (b), but concentrating on just the low-mass clusters, $M_{cluster} < 1000 M_{\Sun}$ with (a) showing a linear plot and (b) showing a log plot.  The dark gray region identifies the MASSCLEAN solution for the range of variation of $M_{max}$.  The location of real clusters from \citeauthor*{weidner2010} \citeyear{weidner2010}, \citeauthor*{weidner2013} \citeyear{weidner2013}, and calculated using MASSCLEAN{\fontfamily{ptm}\selectfont \textit{max}} are shown as green, red, and yellow dots, respectively.  \normalsize}\label{fig:paper5-08}
\end{figure*}

It is with this goal in mind, that we created the newest application in the MASSCLEAN package, MASSCLEAN{\fontfamily{ptm}\selectfont \textit{max}}.  This application uses yet another, newly created MASSCLEAN{\fontfamily{ptm}\selectfont \textit{colors}} database based on $25$ $million$ simulated clusters. The newest version of the MASSCLEAN{\fontfamily{ptm}\selectfont \textit{colors}} database was built using MASSCLEAN (\citeauthor*{paper1} \citeyear{paper1}), Kroupa IMF (\citeauthor*{kroupa2001} \citeyear{kroupa2001}), and Padova stellar evolutionary models with $Z=0.008$ (\citeauthor*{padova2008} \citeyear{padova2008};). Compared to the previous version of the database (\citeauthor*{paper3} \citeyear{paper3}; \citeauthor*{paper4} \citeyear{paper4}), it contains clusters only in the $200-1,000$ $M_{\Sun}$ range, ages in the $6.00-10.13$ $\log(age/yr)$ range, and the mass of the most massive star in the distribution, $M_{max}$, is included. MASSCLEAN{\fontfamily{ptm}\selectfont \textit{max}} then considers the stellar cluster's known mass and age (\citeauthor*{paper4} \citeyear{paper4}), $M_{V}$, $(B-V)_{0}$, and $(U-B)_{0}$ (\citeauthor*{hunter2003} \citeyear{hunter2003}), and using the colors and magnitudes predicted in the database and through probabilistic inference, it finds the most probable value for $M_{max}$ for each cluster. 

$M_{max}$ values computed by MASSCLEAN{\fontfamily{ptm}\selectfont \textit{max}} for 40 low-mass clusters ($M_{cluster}$ in the $200-1,000$ $M_{\Sun}$ range) are presented in Figure \ref{fig:paper5-08} as yellow dots and listed in the Table \ref{table2}. The 40 MASSCLEAN clusters were selected from \citeauthor*{paper4} \citeyearpar{paper4} catalog to be young and low mass, so their most massive star will still be present in the mass distribution. The \citeauthor*{weidner2010} \citeyearpar{weidner2010} clusters, with their maximum mass as derived by them, are presented as green dots, and the additional clusters from \citeauthor*{weidner2013} \citeyearpar{weidner2013} are presented as red dots (unfortunately, the majority of them do not include error bars). In Figure \ref{fig:paper5-08} (a) is also presented, along with our 40 clusters and the \citeauthor*{weidner2010} \citeyearpar{weidner2010} clusters, the range of variation of $M_{max}$ as the broad gray area. The {\it canonical} forms given by $M_1(M_{limit})$, \citeauthor*{kroupa2011} \citeyearpar{kroupa2011}, and \citeauthor*{weidner2013} \citeyearpar{weidner2013} are presented as blue, cyan, and magenta lines, respectively. All sets of yellow, red, and green clusters show a pretty similar placement within the $M_{max}$ range,  determined in Section \S 4. Figure \ref{fig:paper5-08} (b) is identical to Figure \ref{fig:paper5-08} (a), but shown in logarithmic scale.  Note, these figures are virtually identical to what was shown in Figure \ref{fig:paper5-05} (b), only here we have limited the plot to a smaller stellar and cluster mass-range.  

It is important to remember, the MASSCLEAN{\fontfamily{ptm}\selectfont \textit{max}} method provides only an estimate of the $M_{max}$.  Further, it assumes we have obtained {\it perfect} values for the age and mass of the cluster under study.  However, any age and mass derived for a stellar cluster, will have some error associated with it, no matter what method is used to obtain age and mass.  We believe the accuracy with which we can estimate $M_{max}$ would be increased if we are able to include the error bars in age and mass of the cluster under study in the analysis.  This will be the subject of future work.

\section{An Analytical Description of $M_{max}$ range and $M_{max}-M_{cluster}$ relation}

As described in the previous sections, our MASSCLEAN simulations indicate that the maximum stellar mass in a stellar cluster, $M_{max}$, covers a range of mass that is dependent on the mass of the cluster. The well-behaved shape of this range presented in Figures \ref{fig:paper5-04} and \ref{fig:paper5-05} and shaded grey, lends itself to finding an analytical form for the upper and lower limits. The logarithmic plot from Figure \ref{fig:paper5-05} (b) shows that a power law is a good fit for both limits in the case of {\it unsaturated} clusters.

\begin{figure*}[htp]
\centering
\subfigure[]{\includegraphics[angle=270,width=0.48\textwidth]{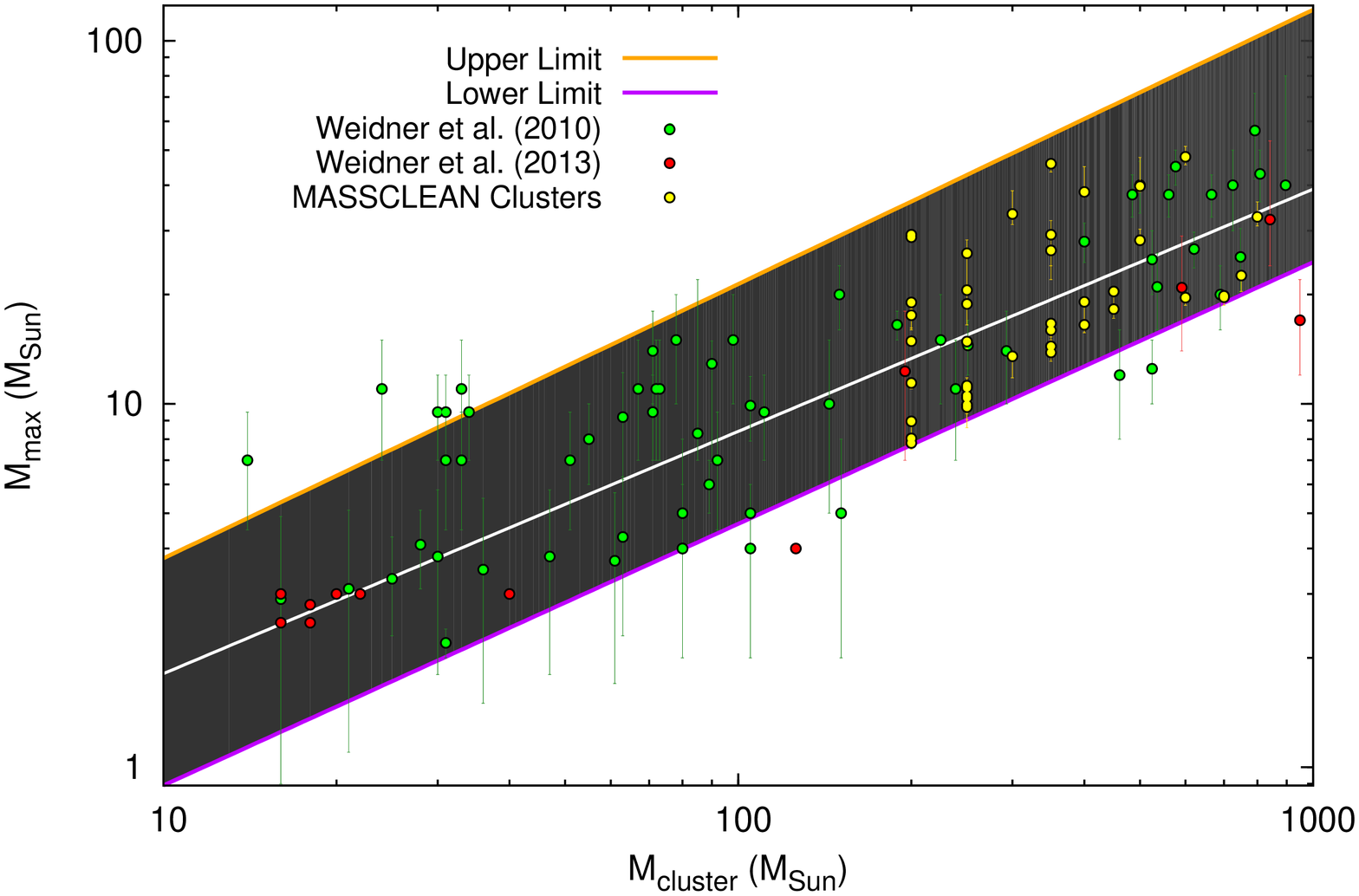}} 
\subfigure[]{\includegraphics[angle=270,width=0.48\textwidth]{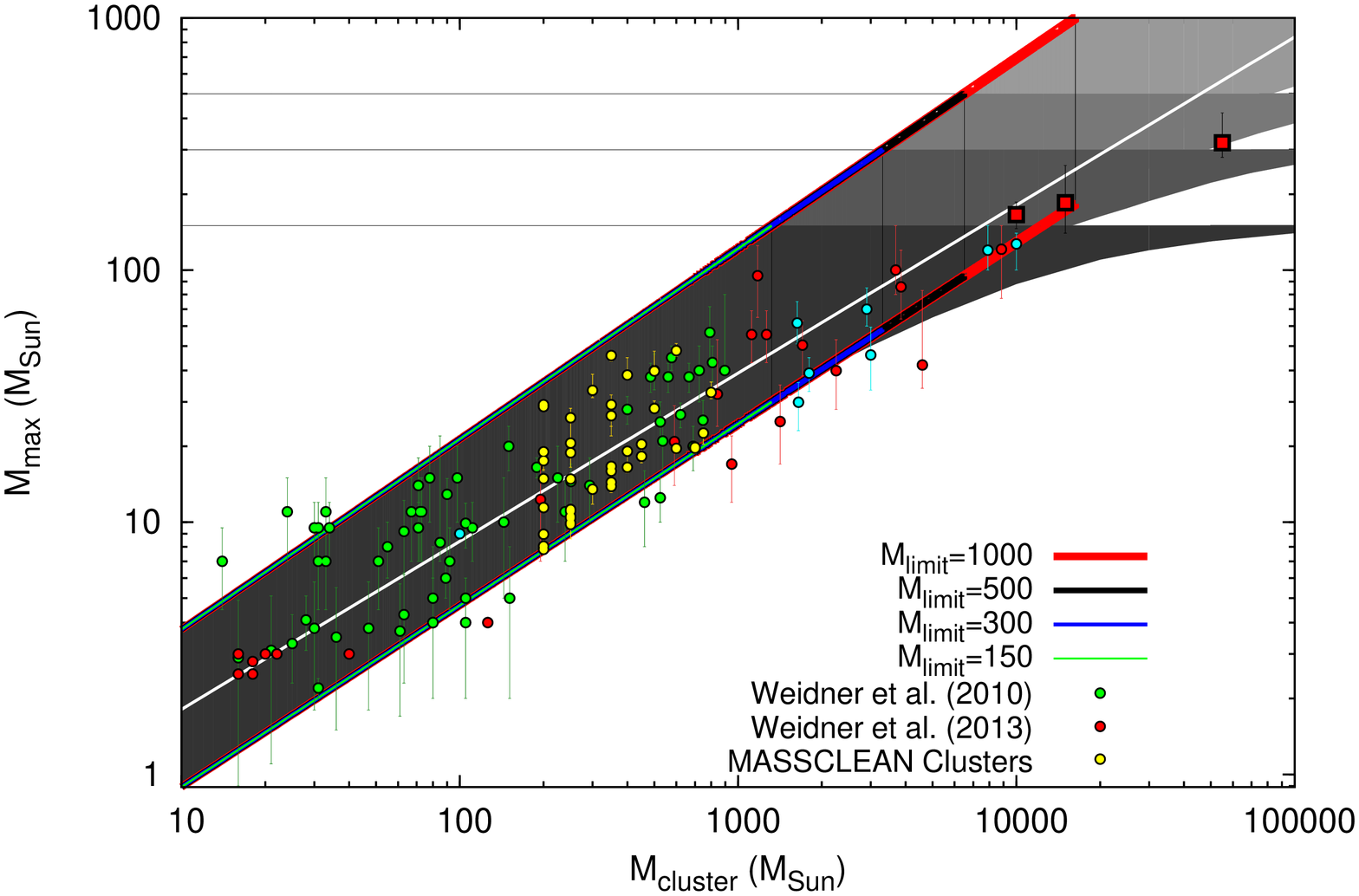}} 
\caption[]{\small (a) the same as Figure 8 (b), but showing our analytical fit to the grey regions upper and lower limits.  The white line presents results from hydrodynamical simulations. The location of real clusters from \citeauthor*{weidner2010} \citeyear{weidner2010}, \citeauthor*{weidner2013} \citeyear{weidner2013}, and calculated using MASSCLEAN{\fontfamily{ptm}\selectfont \textit{max}} are shown as green, red, and yellow dots, respectively. (b)  Similar to (a), but extends to a larger cluster mass, and includes the effect of applying a upper mass limit, $M_{limit}$. \citeauthor*{crowther2010} \citeyear{crowther2010} clusters are presented as red squares, and additional clusters from literature are shows as cyan dots. \normalsize}\label{fig:paper5-09}
\end{figure*}

{\sc Optimal Sampling} (\citeauthor*{kroupa2011} \citeyear{kroupa2011}) uses a single valued $M_{max}-M_{cluster}$ relation, computed with a $150$ $M_{\Sun}$ {\it canonical limit}, and described in a complicated form by Equation (\ref{eq:24}). On the other hand, hydrodynamical simulations (e.g. \citeauthor*{bonnell2004} \citeyear{bonnell2004}; \citeauthor*{peters2010} \citeyear{peters2010}, \citeyear{peters2011}; \citeauthor*{kroupa2011} \citeyear{kroupa2011}) give the dependence between the mass of the most massive star and cluster's mass in the power law form:

\begin{equation} \label{eq:27}
   M_{max} =0.39 M_{cluster}^{2/3}
\end{equation}

This form corresponds to the best fit to these hydrodyamnical simulations, though some scatter might still exist around it (e.g. \citeauthor*{bonnell2004} \citeyear{bonnell2004}).

From our simulations, we have determined that the $M_{max}-M_{cluster}$ relation is not single valued. The range of variation is best described by an upper ($M_{max_{1}}$) and lower limit ($M_{max_{2}}$). These two limits could be written as a power law in a similar form to Equation (\ref{eq:27}) as:

\begin{equation} \label{eq:28}
   M_{max_{1,2}} = k_{1,2} M_{cluster}^{\beta_{1,2}}
\end{equation}
with: 
$k_{1}=0.66$,  
$k_{2}=0.17$,  
$\beta_{1}=0.755$, and 
$\beta_{2}=0.720$ .

The upper and lower limits given by the Equation (\ref{eq:28}) are presented if Figure \ref{fig:paper5-09} (a) as the orange and purple lines, respectively. The $M_{max}-M_{cluster}$ relation determined from hydrodynamical simulations (Equation (\ref{eq:27})) is presented as the white line. The range of $M_{max}$ variation is presented as the gray area. The MASSCLEAN clusters, \citeauthor*{weidner2010} \citeyearpar{weidner2010} clusters, and \citeauthor*{weidner2013} \citeyearpar{weidner2013} are presented as yellow, green, and red dots, respectively, similarly to the Figure \ref{fig:paper5-08} (b).

In the Figure \ref{fig:paper5-09} (b) we extended both the $M_{cluster}$ and $M_{max}$ range.  $M_{max_{1,2}}$ are presented as green, blue, black, and red lines for $M_{limit}=150$ $M_{\Sun}$, $300$ $M_{\Sun}$, $500$ $M_{\Sun}$, and $1,000$ $M_{\Sun}$, respectively. We note that $M_{max_{1,2}}$ limits given by the Equation (\ref{eq:28}) stand for {\it unsaturated} clusters, regardless of the $M_{limit}$. In addition to MASSCLEAN clusters, \citeauthor*{weidner2010} \citeyearpar{weidner2010} clusters, and \citeauthor*{weidner2013} \citeyearpar{weidner2013} clusters, we also show \citeauthor*{crowther2010} \citeyearpar{crowther2010} clusters as red squares, and other clusters from literature as cyan dots.

Based on the fact that $M_{max}$ is related to the $\alpha_{3}$ value (Equations (\ref{eq:17}), (\ref{eq:20}) and  (\ref{eq:22})), just to emphasize this dependence Equation (\ref{eq:28}) could be rewritten as:

\begin{equation} \label{eq:29}
   M_{max_{1,2}} = k_{1,2} M_{cluster}^{\frac{1}{\alpha_{3}-1 \mp \delta_{1,2}} }
\end{equation}
with: 
$\delta_{1}=0.0255$, 
$\delta_{2}=0.0388$, and $\alpha_{3}=2.35$.

Regardless of the way the dependence is written, we find that the $M_{max}-M_{cluster}$ relation for {\it unsaturated} clusters is described by the upper and lower limits $M_{max_{1,2}}$. Both of these values correspond to power laws of the mass of the cluster, and are independent of any particular value for $M_{limit}$ when applied to unsaturated, low-mass stellar clusters.

\section{Summary and Conclusion}

The IMF is one of the most fundamental of astrophysical distribution functions, with broad applicability. Traditionally, the IMF functional form was predominately used to fit observational data. However in current investigations, simulated data is becoming increasingly important. This warrants a careful consideration that the methods used for filling an IMF distribution in such applications are done correctly. The most often used (and convenient) way to do this, through random sampling, does not properly fill an IMF distribution, as described by \citeauthor*{kroupa2011} \citeyearpar{kroupa2011}. 

{\sc Optimal Sampling} (\citeauthor*{kroupa2011} \citeyear{kroupa2011}) does fill the IMF correctly. But because it allows for multiple stars per mass bin, it lacks a certain {\it resolution} to pursue additional fundamental questions about the distribution.  
 Moreover, it leads to a single-valued $M_{max}-M_{cluster}$ relation.   \citeauthor*{kroupa2011} \citeyearpar{kroupa2011} attempt to explain the range observed in this relationship (our Figures \ref{fig:paper5-05} (b), \ref{fig:paper5-08} and \ref{fig:paper5-09}, and their Figures 4-5) as due to observational error and stochastic variations in the intrinsic pre-cluster cloud conditions, which may vary a single valued $M_{max}-M_{cluster}$ relation.
We presented our {\bf M}ASSCLEAN {\bf IMF S}ampling (MIMFS) method, which is able to properly fill the IMF using just one star per mass bin and requires no assumption about the $M_{max}-M_{cluster}$ relation or $M_{limit}$. 

From our {\it 10 million} MASSCLEAN Monte Carlo simulations we determined the expected mass range of the most massive star in the stellar mass distribution as a function of the mass of the cluster.   As a check on these simulations, we confirm this maximum mass range is independent of $M_{limit}$ for {\it unsaturated} clusters. 
What is particularly validating about our method is it {\it predicts} the $M_{max}-M_{cluster}$ relation to be a range, not single-valued.  It even predicts the correct upper and lower mass range when compared to real clusters.  

We described our method to determine $M_{max}$ from $U,B,V$ integrated colors and magnitudes using {\it 25 million} MASSCLEAN Monte Carlo simulations.   With it, we estimate the maximum stellar mass for 40 LMC clusters.  These values of maximum stellar mass, relative to the cluster total mass, are consistent with previous determinations using different methods on other similar-mass stellar clusters (e.g. \citeauthor*{weidner2010} \citeyear{weidner2010}). 

Finally, we provided an analytical, power-law, description of the $M_{max}$ range, which enabled us to cleanly describe the $M_{max}-M_{cluster}$ relation. For {\it unsaturated cluters}, the $M_{max}-M_{cluster}$ relation is not single-valued. It is described by an {\it upper} and {\it lower} limit. Both of these limits correspond to power-law functions of the stellar cluster mass, $M_{cluster}$, and are independent of $M_{limit}$ for {\it unsaturated} clusters.  Such a relationship is consistent with previous results from hydrodynamical simulations. 

\acknowledgements
We thank the referee for useful comments and suggestions. 
We are grateful to suggestions made to an early draft of this work by Bruce Elmegreen and Soeren Larsen.  Their ideas lead to significant improvements in the presentation. 
This material is based upon work supported by the National Science Foundation under Grant No. 0607497 and 1009550, to the University of Cincinnati. BP acknowledges additional support from CMP. 

\vskip 1cm

{\bf Appendix}

\appendix



\begin{deluxetable}{lrrc}
\tablecolumns{9}
\tablewidth{0pt}
\tabletypesize{\scriptsize}
\tablecaption{Clusters with $M_{max}$ Values \label{table1}}
\tablehead{
\colhead{{\scriptsize Name}} & \colhead{{\scriptsize$M$}} & \colhead{{\scriptsize$M_{max}$}} & \colhead{{\scriptsize Reference\tablenotemark{a}}}  \\
\colhead{{\scriptsize }} & \colhead{{\scriptsize$(M_{\Sun})$}} & \colhead{{\scriptsize$(M_{\Sun})$}} & \colhead{{\scriptsize }}  \\
\colhead{{\tiny$1$}} &  \colhead{{\tiny$2$}} &  \colhead{{\tiny$3$}} & \colhead{{\tiny$4$}}   

}
\startdata
\\
{\tiny NGC 3603} &  {\tiny$10000$ \phn \phn \phn} & {\tiny$166^{+20 \phn \phn}_{-20} $ } & {\tiny 1 } \\

{\tiny Arches} &  {\tiny$15000$ \phn \phn \phn} & {\tiny$185^{+75 \phn \phn}_{-45} $ } & {\tiny 1, 2 } \\

{\tiny R136} &  {\tiny$55000$ \phn \phn \phn} & {\tiny$320^{+100 \phn}_{-40} $ } & {\tiny 1 } \\

\\
\cline{1-4} \\
{\tiny IRAS 05274+3345} &  {\tiny$14^{+15 \phd }_{-7}$ } & {\tiny$7.0\pm2.5 \phd \phd$ } & {\tiny 3 } \\

{\tiny Mol 139} &  {\tiny$16\pm8$ \phd \phd } & {\tiny$2.9\pm2.0 \phd \phd$ } & {\tiny 3 } \\

{\tiny Mol 143} &  {\tiny$21\pm10  $ } & {\tiny$3.1\pm2.0$ \phn \phd} & {\tiny 3 } \\

{\tiny IRAS 06308+0402} &  {\tiny$24^{+25}_{-13} \phn $ } & {\tiny$11.0\pm4.0$ \phn \phd} & {\tiny 3 } \\

{\tiny VV Ser} &  {\tiny$25^{+27}_{-13} \phn $ } & {\tiny$3.3\pm1.0$ \phn \phd} & {\tiny 3 } \\

{\tiny VY Mon} &  {\tiny$28^{+29}_{-15} \phn $ } & {\tiny$4.1\pm1.0$ \phn \phd} & {\tiny 3 } \\

{\tiny Mol 8A} &  {\tiny$30\pm15  $ } & {\tiny$3.8\pm2.0$ \phn \phd} & {\tiny 3 } \\

{\tiny IRAS 05377+3548} &  {\tiny$30^{+32}_{-15} \phn $ } & {\tiny$9.5\pm2.5$ \phn \phd} & {\tiny 3 } \\

{\tiny Ser SVS2} &  {\tiny$31^{+31}_{-16} \phn $ } & {\tiny$2.2\pm0.2$ \phn \phd} & {\tiny 3 } \\

{\tiny Tau-Aur} &  {\tiny$31^{+36}_{-18} \phn $ } & {\tiny$7.0\pm2.5$ \phn \phd} & {\tiny 3 } \\

{\tiny IRAS 05553+1631} &  {\tiny$31^{+33}_{-16} \phn $ } & {\tiny$9.5\pm2.5$ \phn \phd} & {\tiny 3 } \\

{\tiny IRAS 05490+2658} &  {\tiny$33^{+36}_{-17} \phn $ } & {\tiny$7.0\pm2.5$ \phn \phd} & {\tiny 3 } \\

{\tiny IRAS 03064+5638} &  {\tiny$33^{+36}_{-17} \phn $ } & {\tiny$11.0\pm4.0$ \phn \phd} & {\tiny 3 } \\

{\tiny IRAS 06155+2319} &  {\tiny$34^{+35}_{-18} \phn $ } & {\tiny$9.5\pm2.5$ \phn \phd} & {\tiny 3 } \\

{\tiny Mol 50} &  {\tiny$36\pm18 $ } & {\tiny$3.5\pm2.0$ \phn \phd} & {\tiny 3 } \\

{\tiny Mol 11} &  {\tiny$47\pm20 $ } & {\tiny$3.8\pm2.0$ \phn \phd} & {\tiny 3 } \\

{\tiny IRAS 06058+2138} &  {\tiny$51^{+54}_{-27} \phn $ } & {\tiny$7.0\pm2.5$ \phn \phd} & {\tiny 3 } \\

{\tiny NGC 2023} &  {\tiny$55^{+58}_{-28} \phn $ } & {\tiny$8.0\pm2.0$ \phn \phd} & {\tiny 3 } \\

{\tiny Mol 3} &  {\tiny$61\pm20 $ } & {\tiny$3.7\pm2.0$ \phn \phd} & {\tiny 3 } \\

{\tiny Mol 160} &  {\tiny$63\pm20 $ } & {\tiny$4.3\pm2.0$ \phn \phd} & {\tiny 3 } \\

{\tiny NGC 7129} &  {\tiny$63^{+104}_{-33} $ } & {\tiny$9.2\pm3.0$ \phn \phd} & {\tiny 3 } \\

{\tiny IRAS 06068+20303} &  {\tiny$67^{+70}_{-35} \phn $ } & {\tiny$11.0\pm4.0$ \phn \phd} & {\tiny 3 } \\

{\tiny IRAS 00494+5617} &  {\tiny$71^{+74}_{-37} \phn $ } & {\tiny$9.5\pm2.5$ \phn \phd} & {\tiny 3 } \\

{\tiny V921 Sco} &  {\tiny$71^{+429}_{-36} $ } & {\tiny$14.0\pm4.0$ \phn \phd} & {\tiny 3 } \\

{\tiny IRAS 05197+3355} &  {\tiny$72^{+75}_{-38} \phn $ } & {\tiny$11.0\pm4.0$ \phn \phd} & {\tiny 3 } \\

{\tiny IRAS 05375+3540} &  {\tiny$73^{+78}_{-38} \phn $ } & {\tiny$11.0\pm4.0$ \phn \phd} & {\tiny 3 } \\

{\tiny IRAS 02593+6016} &  {\tiny$78^{+81}_{-41} \phn $ } & {\tiny$15.0\pm5.0$ \phn \phd} & {\tiny 3 } \\

{\tiny Cha I} &  {\tiny$80^{+91}_{-46} \phn $ } & {\tiny$5.0\pm3.0$ \phn \phd} & {\tiny 3 } \\

{\tiny Mol 103} &  {\tiny$80\pm 20 $ } & {\tiny$4.0\pm2.0$ \phn \phd} & {\tiny 3 } \\

{\tiny NGC 2071} &  {\tiny$80^{+89}_{-44} \phn $ } & {\tiny$4.0\pm2.0$ \phn \phd} & {\tiny 3 } \\

{\tiny MWC 297} &  {\tiny$85\pm 60 $ } & {\tiny$8.3^{+13.7\phn \phd}_{-1.3}$} & {\tiny 3 } \\

{\tiny IC 348} &  {\tiny$89^{+92}_{-46} \phn $ } & {\tiny$6.0\pm1.0$ \phn \phd} & {\tiny 3 } \\

{\tiny BD 40$^{\circ}$ 4124} &  {\tiny$90^{+106}_{-49} $ } & {\tiny$12.9^{+2.0\phn \phd \phd}_{-6.0}$} & {\tiny 3 } \\

{\tiny IRAS 06056+2131} &  {\tiny$92^{+97}_{-49} \phn $ } & {\tiny$7.0\pm2.5$ \phn \phd} & {\tiny 3 } \\

{\tiny IRAS 05100+3723} &  {\tiny$98^{+103}_{-51}  $ } & {\tiny$15.0\pm5.0$ \phn \phd} & {\tiny 3 } \\

{\tiny R CrA} &  {\tiny$105^{+114}_{-55}  $ } & {\tiny$4.0\pm2.0$ \phn \phd} & {\tiny 3 } \\

{\tiny NGC 1333} &  {\tiny$105^{+111}_{-54}  $ } & {\tiny$5.0\pm1.0$ \phn \phd} & {\tiny 3 } \\

{\tiny Mol 28} &  {\tiny$105\pm 20  $ } & {\tiny$9.9\pm2.0$ \phn \phd} & {\tiny 3 } \\

{\tiny IRAS 02575+6017} &  {\tiny$111^{+116}_{-57}  $ } & {\tiny$9.5\pm2.5$ \phn \phd} & {\tiny 3 } \\

{\tiny W40} &  {\tiny$144^{+576}_{-80}  $ } & {\tiny$10.0\pm5.0$ \phn \phd} & {\tiny 3 } \\

{\tiny $\sigma$ Ori} &  {\tiny$150^{+155}_{-76}  $ } & {\tiny$20.0\pm4.0$ \phn \phd} & {\tiny 3 } \\

{\tiny NGC 2068} &  {\tiny$151^{+169}_{-86}  $ } & {\tiny$5.0\pm3.0$ \phn \phd} & {\tiny 3 } \\

{\tiny NGC 2384} &  {\tiny$189^{+192}_{-95}  $ } & {\tiny$16.5\pm1.5$ \phn \phd} & {\tiny 3 } \\

{\tiny Mon R2} &  {\tiny$225^{+236}_{-117}  $ } & {\tiny$15.0\pm5.0$ \phn \phd} & {\tiny 3 } \\

{\tiny IRAS 06073+1249} &  {\tiny$239^{+242}_{-120}  $ } & {\tiny$11.0\pm4.0$ \phn \phd} & {\tiny 3 } \\

{\tiny Trumpler 24} &  {\tiny$251^{+291}_{-131}  $ } & {\tiny$14.5\pm2.5$ \phn \phd} & {\tiny 3 } \\

{\tiny IC 5146} &  {\tiny$293^{+305}_{-226}  $ } & {\tiny$14.0\pm4.0$ \phn \phd} & {\tiny 3 } \\

{\tiny HD 52266} &  {\tiny  $400\pm 350  $ } & {\tiny$28.0\pm3.5$ \phn \phd} & {\tiny 3 } \\

{\tiny HD 57682} &  {\tiny$400\pm 350  $ } & {\tiny$28.0\pm3.5$ \phn \phd} & {\tiny 3 } \\

{\tiny Alicante 5} &  {\tiny$461^{+516}_{-234}  $ } & {\tiny$12.0\pm4.0$ \phn \phd} & {\tiny 3 } \\

{\tiny Cep OB3b} &  {\tiny$485^{+497}_{-243}  $ } & {\tiny$37.7\pm5.0$ \phn \phd} & {\tiny 3 } \\

{\tiny HD 153426} &  {\tiny$500\pm 350  $ } & {\tiny$40.0\pm6.5$ \phn \phd} & {\tiny 3 } \\

{\tiny NGC 2264} &  {\tiny$525^{+537}_{-267}  $ } & {\tiny$25.0\pm5.0$ \phn \phd} & {\tiny 3 } \\

{\tiny Sh2-294} &  {\tiny$525^{+540}_{-267}  $ } & {\tiny$12.5\pm2.5$ \phn \phd} & {\tiny 3 } \\

{\tiny RCW 116B} &  {\tiny$536^{+557}_{-276}  $ } & {\tiny$21.0\pm5.0$ \phn \phd} & {\tiny 3 } \\

{\tiny NGC 6383} &  {\tiny$561^{+563}_{-281}  $ } & {\tiny$37.7\pm5.0$ \phn \phd} & {\tiny 3 } \\

{\tiny Alicante 1} &  {\tiny$577^{+583}_{-290}  $ } & {\tiny$45.0\pm5.0$ \phn \phd} & {\tiny 3 } \\

{\tiny HD 52533} &  {\tiny$621^{+1077}_{-417}  $ } & {\tiny$26.7\pm3.0$ \phn \phd} & {\tiny 3 } \\

{\tiny Sh2-128} &  {\tiny$666^{+736}_{-342}  $ } & {\tiny$37.7\pm5.0$ \phn \phd} & {\tiny 3 } \\

{\tiny NGC 2024} &  {\tiny$690^{+706}_{-350}  $ } & {\tiny$20.0\pm4.0$ \phn \phd} & {\tiny 3 } \\

{\tiny HD 195592} &  {\tiny$725^{+757}_{-364}  $ } & {\tiny$40.0\pm10.0$ \phd} & {\tiny 3 } \\

{\tiny Sh2-173} &  {\tiny$748^{+901}_{-395}  $ } & {\tiny$25.4\pm5.0$ \phn \phd} & {\tiny 3 } \\

{\tiny DBSB 48} &  {\tiny$792^{+1126}_{-416}  $ } & {\tiny$56.6\pm15.0$ \phd} & {\tiny 3 } \\

{\tiny NGC 2362} &  {\tiny$809^{+823}_{-409}  $ } & {\tiny$43.0\pm7.0$ \phn \phd} & {\tiny 3 } \\

{\tiny Pismis 11} &  {\tiny$896^{+938}_{-448}$ } & {\tiny$40.0^{+40.0\phn}_{-0.0}$ } & {\tiny 3 } \\

\\
\cline{1-4} \\

{\tiny Taurus-Auriga 5} &  {\tiny$16 \phn \phn \phn$ } & {\tiny$2.5 \phn \phn \phn$ } & {\tiny 4 } \\

{\tiny Taurus-Auriga 2} &  {\tiny$16 \phn \phn \phn$ } & {\tiny$3.0 \phn \phn \phn$ } & {\tiny 4 } \\

{\tiny Taurus-Auriga 4} &  {\tiny$18 \phn \phn \phn$ } & {\tiny$2.5 \phn \phn \phn$ } & {\tiny 4 } \\

{\tiny Lupus 3} &  {\tiny$18 \phn \phn \phn$ } & {\tiny$2.8 \phn \phn \phn$ } & {\tiny 4 } \\

{\tiny Cha I 2} &  {\tiny$20 \phn \phn \phn$ } & {\tiny$3.0 \phn \phn \phn$ } & {\tiny 4 } \\

{\tiny IC 348 1} &  {\tiny$126 \phn \phn \phn$ } & {\tiny$4.0 \phn \phn \phn$ } & {\tiny 4 } \\

{\tiny LkH$\alpha$ 101} &  {\tiny$195^{+295 \phn}_{-123}$ } & {\tiny$12.3^{+5.7\phn \phn}_{-5.3}$ } & {\tiny 4 } \\

{\tiny RCW 36} &  {\tiny$591^{+619 \phn}_{-305}$ } & {\tiny$20.9^{+8.1\phn \phn}_{-6.9}$ } & {\tiny 4 } \\

{\tiny [BDSB2003] 164} &  {\tiny$842^{+1065 }_{-429}$ } & {\tiny$32.2^{+20.8 \phn}_{-8.2}$ } & {\tiny 4 } \\

{\tiny [FSR2007] 777} &  {\tiny$949^{+2166 }_{-758}$ } & {\tiny$17.0\pm 5.0 \phn$ } & {\tiny 4 } \\

{\tiny NGC 6530} &  {\tiny$1118^{+1132}_{-564}$ } & {\tiny$55.5^{+13.5\phn}_{-12.5}$ } & {\tiny 4 } \\

{\tiny [FSR2007] 734} &  {\tiny$1175^{+1202}_{-833}$ } & {\tiny$95.0\pm30.0$ } & {\tiny 4 } \\

{\tiny [DBSB2003] 177} &  {\tiny$1265^{+1266}_{-633}$ } & {\tiny$55.5^{+13.5\phn}_{-12.5}$ } & {\tiny 4 } \\

{\tiny [DB2000] 52} &  {\tiny$1416^{+1591}_{-724}$ } & {\tiny$25.1^{+9.9\phn \phn}_{-8.1}$ } & {\tiny 4 } \\

{\tiny [DB2000] 26} &  {\tiny$1705^{+1721}_{-852}$ } & {\tiny$50.4^{+12.6 \phn}_{-12.4}$ } & {\tiny 4 } \\

{\tiny RCW 38} &  {\tiny$2251^{+2276}_{-1132}$ } & {\tiny$39.9^{+13.1 \phn}_{-11.9}$ } & {\tiny 4 } \\

{\tiny Mercer 23} &  {\tiny$3687^{+3793}_{-1859}$ } & {\tiny$100.0^{+50.0 \phn}_{-20.0}$ } & {\tiny 4 } \\

{\tiny NGC 2103} &  {\tiny$3853^{+3905}_{-1937}$ } & {\tiny$85.8^{+34.2 \phn}_{-21.8}$ } & {\tiny 4 } \\

{\tiny NGC 6231} &  {\tiny$4595^{+4676}_{-2312}$ } & {\tiny$42.0^{+41.0 \phn}_{-8.0}$ } & {\tiny 4 } \\

{\tiny Westerlund 2} &  {\tiny$8845^{+9009}_{-4456}$ } & {\tiny$121.0^{+29.0\phn}_{-43.8}$ } & {\tiny 4 } \\

\\
\cline{1-4} \\

{\tiny Danks 2} &  {\tiny$2900 \phn \phn \phn$ } & {\tiny$70^{+15\phn \phn \phn}_{-10}$ } & {\tiny 5, 6 } \\

{\tiny Danks 1} &  {\tiny$7900 \phn \phn \phn$ } & {\tiny$120^{+30\phn \phn \phn}_{-20}$ } & {\tiny 5, 6 } \\

{\tiny RCW 79} &  {\tiny$3000 \phn \phn \phn$ } & {\tiny$46.1^{+13.3 }_{-12.7}$ } & {\tiny 5, 7 } \\

{\tiny Trumpler 14} &  {\tiny$10000 \phn \phn \phn$ } & {\tiny$127^{+13\phn \phn \phn}_{-27}$ } & {\tiny 8, 9 } \\

{\tiny $\rho$ Oph} &  {\tiny$100 \phn \phn \phn$ } & {\tiny$9\phn \phn \phn \phn \phn \phn \phn$ } & {\tiny 10 } \\

{\tiny ONC} &  {\tiny$1800 \phn \phn \phn$ } & {\tiny$39\pm6\phn \phn \phd \phd$ } & {\tiny 10 } \\

{\tiny NGC 6611} &  {\tiny$1630 \phn \phn \phn$ } & {\tiny$61^{+14\phn \phn \phn}_{-10}$ } & {\tiny 11, 4 } \\

{\tiny RCW 120} &  {\tiny$1650 \phn \phn \phn$ } & {\tiny$29.9^{+5.9 \phn}_{-6.8}$ } & {\tiny 7, 12 }

\enddata
\tablenotetext{a}{{\scriptsize References: {\bf 1.} \citeauthor*{crowther2010} \citeyearpar{crowther2010}; {\bf 2.} \citeauthor*{clarkson2012} \citeyearpar{clarkson2012}; {\bf 3.} \citeauthor*{weidner2010} \citeyearpar{weidner2010}; {\bf 4.} \citeauthor*{weidner2013} \citeyearpar{weidner2013}; {\bf 5.} \citeauthor*{chene2012} \citeyearpar{chene2012}; {\bf 6.} \citeauthor*{davies2012} \citeyearpar{davies2012}; {\bf 7.} \citeauthor*{martins2010} \citeyearpar{martins2010}; {\bf 8.} \citeauthor*{ascenso2007} \citeyearpar{ascenso2007} ; {\bf 9.}  \citeauthor*{hur2012} \citeyearpar{hur2012} ; {\bf 10.} \citeauthor*{crowther2012} \citeyearpar{crowther2012}; {\bf 11.} \citeauthor*{bonatto2006} \citeyearpar{bonatto2006}; {\bf 12.} \citeauthor*{deharveng2009} \citeyearpar{deharveng2009}. }}
\end{deluxetable}

\begin{deluxetable}{lrrrrrrrr}
\tablecolumns{9}
\tablewidth{0pt}
\tabletypesize{\scriptsize}
\tablecaption{LMC Clusters with $M_{max}$ Values \label{table2}}
\tablehead{
\colhead{}    &  \multicolumn{4}{c}{{\scriptsize Integrated Photometry (Hunter et al. 2003)}} & \colhead{{\tiny }}  &   \multicolumn{3}{c}{{\scriptsize MASSCLEAN}} \\
\cline{2-5} \cline{7-9} \\
\colhead{{\scriptsize Name(s)}} & \colhead{{\scriptsize$M_{V}$}} & \colhead{{\scriptsize$(U-B)_{0}$}} & \colhead{{\scriptsize$(B-V)_{0}$}} & \colhead{{\scriptsize$(V-R)_{0}$}}  & \colhead{{\tiny }}  & \colhead{{\scriptsize Age\tablenotemark{a}}}  & \colhead{{\scriptsize Mass\tablenotemark{a}}} & \colhead{{\scriptsize $M_{max}$\tablenotemark{b}}} \\
\colhead{{\scriptsize }} & \colhead{{\scriptsize$(mag)$}} & \colhead{{\scriptsize$(mag)$}} & \colhead{{\scriptsize$(mag)$}} & \colhead{{\scriptsize$(mag)$}}   & \colhead{{\tiny }} & \colhead{{\scriptsize $(log)$}}   &  \colhead{{\scriptsize $(M_{\Sun})$}} & \colhead{{\scriptsize $(M_{\Sun})$}}   \\
\colhead{{\tiny$1$}} &  \colhead{{\tiny$2$}} & \colhead{{\tiny$3$}} & \colhead{{\tiny$4$}} & \colhead{{\tiny$5$}} & \colhead{{\tiny }}  & \colhead{{\tiny$6$}} & \colhead{{\tiny$7$}} & \colhead{{\tiny$8$}}

}
\startdata
{\tiny BSDL2208} & {\tiny$-5.618\pm 0.012$  \phn} & {\tiny$-0.894\pm 0.005  $  \phn} & {\tiny$-0.296\pm 0.012  $  \phn} & {\tiny$-0.086\pm 0.026  $  \phn} &  {\tiny} & {\tiny$6.73 $ } & {\tiny$200 \phn$ } & {\tiny$29.24^{+0.76 \phn}_{-0.74} $} \\
{\tiny NGC1837.SL217} & {\tiny$-7.223\pm 0.004$  \phn} & {\tiny$-0.884\pm 0.002  $  \phn} & {\tiny$-0.054\pm 0.004  $  \phn} & {\tiny$-0.036\pm 0.009  $  \phn} &  {\tiny} & {\tiny$6.81   $ } & {\tiny$200 \phn$ } & {\tiny$28.76\pm0.01 $} \\
{\tiny KMHK263} & {\tiny$-4.671\pm 0.021$  \phn} & {\tiny$-1.079\pm 0.006  $  \phn} & {\tiny$-0.191\pm 0.021  $  \phn} & {\tiny$-0.167\pm 0.057  $  \phn} &  {\tiny} & {\tiny$6.88   $ } & {\tiny$200 \phn$ } & {\tiny$19.02^{+0.38 \phn}_{-3.02} $} \\
{\tiny BSDL579} & {\tiny$-5.300\pm 0.011$  \phn} & {\tiny$-1.025\pm 0.004  $  \phn} & {\tiny$-0.161\pm 0.011  $  \phn} & {\tiny$-0.088\pm 0.027  $  \phn} &  {\tiny} & {\tiny$7.03   $ } & {\tiny$200 \phn$ } & {\tiny$17.54^{+0.71 \phn}_{-1.34} $} \\
{\tiny KMHK612} & {\tiny$-6.634\pm 0.005$  \phn} & {\tiny$1.137\pm 0.005  $  \phn} & {\tiny$1.232\pm 0.007  $  \phn} & {\tiny$0.585\pm 0.008  $  \phn} &  {\tiny} & {\tiny$7.16   $ } & {\tiny$200 \phn$ } & {\tiny$14.88^{+3.37 \phn}_{-0.18} $} \\
{\tiny BSDL2119} & {\tiny$-6.023\pm 0.008$  \phn} & {\tiny$0.867\pm 0.011  $  \phn} & {\tiny$1.566\pm 0.013  $  \phn} & {\tiny$0.875\pm 0.011  $  \phn} &  {\tiny} & {\tiny$7.36   $ } & {\tiny$200 \phn$ } & {\tiny$11.43^{+0.07 \phn}_{-0.43} $} \\
{\tiny SL294.KMHK627} & {\tiny$-5.914\pm 0.032$  \phn} & {\tiny$0.381\pm 0.024  $  \phn} & {\tiny$0.649\pm 0.037  $  \phn} & {\tiny$0.182\pm 0.060  $  \phn} &  {\tiny} & {\tiny$7.53   $ } & {\tiny$200 \phn$ } & {\tiny$8.96^{+0.14 \phn}_{-0.11} $} \\
{\tiny BSDL499} & {\tiny$-3.761\pm 0.039$  \phn} & {\tiny$-0.749\pm 0.011  $  \phn} & {\tiny$-0.169\pm 0.039  $  \phn} & {\tiny$-0.186\pm 0.105  $  \phn} &  {\tiny} & {\tiny$7.54   $ } & {\tiny$200 \phn$ } & {\tiny$7.80^{+0.90 \phn}_{-0.30} $} \\
{\tiny OGLE-LMC0297} & {\tiny$-4.894\pm 0.027$  \phn} & {\tiny$0.264\pm 0.043  $  \phn} & {\tiny$1.198\pm 0.047  $  \phn} & {\tiny$0.627\pm 0.039  $  \phn} &  {\tiny} & {\tiny$7.61   $ } & {\tiny$200 \phn$ } & {\tiny$8.04^{+0.26 \phn}_{-0.05} $} \\
{\tiny BSDL137} & {\tiny$-5.099\pm 0.014$  \phn} & {\tiny$-1.223\pm 0.004  $  \phn} & {\tiny$-0.270\pm 0.014  $  \phn} & {\tiny$-0.080\pm 0.033  $  \phn} &  {\tiny} & {\tiny$6.68   $ } & {\tiny$250 \phn$ } & {\tiny$26.00^{+2.30 \phn}_{-6.00} $} \\
{\tiny BSDL2215} & {\tiny$-5.718\pm 0.010$  \phn} & {\tiny$-0.991\pm 0.003  $  \phn} & {\tiny$-0.198\pm 0.010  $  \phn} & {\tiny$-0.083\pm 0.024  $  \phn} &  {\tiny} & {\tiny$6.95   $ } & {\tiny$250 \phn$ } & {\tiny$20.58^{+0.62 \phn}_{-1.98} $} \\
{\tiny BSDL2704} & {\tiny$-5.473\pm 0.013$  \phn} & {\tiny$-1.029\pm 0.005  $  \phn} & {\tiny$-0.140\pm 0.013  $  \phn} & {\tiny$-0.035\pm 0.028  $  \phn} &  {\tiny} & {\tiny$7.00   $ } & {\tiny$250 \phn$ } & {\tiny$18.85^{+0.35 \phn}_{-2.35} $} \\
{\tiny BSDL358} & {\tiny$-6.681\pm 0.006$  \phn} & {\tiny$0.872\pm 0.007  $  \phn} & {\tiny$1.167\pm 0.009  $  \phn} & {\tiny$0.753\pm 0.008  $  \phn} &  {\tiny} & {\tiny$7.18   $ } & {\tiny$250 \phn$ } & {\tiny$14.84^{+0.06 \phn}_{-0.64} $} \\
{\tiny KMHK900} & {\tiny$-4.448\pm 0.018$  \phn} & {\tiny$-0.883\pm 0.006  $  \phn} & {\tiny$-0.195\pm 0.018  $  \phn} & {\tiny$-0.021\pm 0.047  $  \phn} &  {\tiny} & {\tiny$7.31   $ } & {\tiny$250 \phn$ } & {\tiny$10.54^{+1.26 \phn}_{-0.54} $} \\
{\tiny BSDL1760} & {\tiny$-4.351\pm 0.020$  \phn} & {\tiny$-0.868\pm 0.006  $  \phn} & {\tiny$-0.246\pm 0.020  $  \phn} & {\tiny$-0.045\pm 0.053  $  \phn} &  {\tiny} & {\tiny$7.34   $ } & {\tiny$250 \phn$ } & {\tiny$10.40^{+0.90 \phn}_{-1.00} $} \\
{\tiny OGLE-LMC0169} & {\tiny$-6.335\pm 0.008$  \phn} & {\tiny$0.607\pm 0.008  $  \phn} & {\tiny$0.803\pm 0.011  $  \phn} & {\tiny$0.471\pm 0.013  $  \phn} &  {\tiny} & {\tiny$7.36   $ } & {\tiny$250 \phn$ } & {\tiny$11.22^{+0.18 \phn}_{-0.22} $} \\
{\tiny BSDL917} & {\tiny$-6.279\pm 0.007$  \phn} & {\tiny$0.502\pm 0.005  $  \phn} & {\tiny$0.878\pm 0.008  $  \phn} & {\tiny$0.421\pm 0.012  $  \phn} &  {\tiny} & {\tiny$7.37   $ } & {\tiny$250 \phn$ } & {\tiny$11.07^{+0.23 \phn}_{-0.27} $} \\
{\tiny BSDL256} & {\tiny$-4.329\pm 0.035$  \phn} & {\tiny$-0.811\pm 0.012  $  \phn} & {\tiny$-0.227\pm 0.035  $  \phn} & {\tiny$-0.170\pm 0.087  $  \phn} &  {\tiny} & {\tiny$7.41   $ } & {\tiny$250 \phn$ } & {\tiny$9.80^{+0.52 \phn}_{-0.74} $} \\
{\tiny BSDL25} & {\tiny$-4.267\pm 0.030$  \phn} & {\tiny$-0.795\pm 0.010  $  \phn} & {\tiny$-0.210\pm 0.030  $  \phn} & {\tiny$-0.226\pm 0.079  $  \phn} &  {\tiny} & {\tiny$7.42   $ } & {\tiny$250 \phn$ } & {\tiny$9.94^{+0.26 \phn}_{-1.34} $} \\
{\tiny BSDL295} & {\tiny$-6.303\pm 0.008$  \phn} & {\tiny$-0.927\pm 0.003  $  \phn} & {\tiny$-0.268\pm 0.008  $  \phn} & {\tiny$-0.133\pm 0.018  $  \phn} &  {\tiny} & {\tiny$6.72   $ } & {\tiny$300 \phn$ } & {\tiny$33.34^{+5.26 \phn}_{-2.14} $} \\
{\tiny BSDL2448} & {\tiny$-4.790\pm 0.022$  \phn} & {\tiny$-0.934\pm 0.008  $  \phn} & {\tiny$-0.116\pm 0.022  $  \phn} & {\tiny$-0.001\pm 0.053  $  \phn} &  {\tiny} & {\tiny$7.20   $ } & {\tiny$300 \phn$ } & {\tiny$13.52^{+0.38 \phn}_{-1.72} $} \\
{\tiny KMHK237} & {\tiny$-6.638\pm 0.006$  \phn} & {\tiny$-0.979\pm 0.002  $  \phn} & {\tiny$-0.259\pm 0.006  $  \phn} & {\tiny$-0.143\pm 0.014  $  \phn} &  {\tiny} & {\tiny$6.60   $ } & {\tiny$350 \phn$ } & {\tiny$45.84^{+1.16 \phn}_{-2.34} $} \\
{\tiny BCD1} & {\tiny$-5.786\pm 0.016$  \phn} & {\tiny$-0.893\pm 0.007  $  \phn} & {\tiny$-0.260\pm 0.017  $  \phn} & {\tiny$-0.062\pm 0.032  $  \phn} &  {\tiny} & {\tiny$6.73   $ } & {\tiny$350 \phn$ } & {\tiny$29.23^{+2.77 \phn}_{-7.23} $} \\
{\tiny BSDL2883} & {\tiny$-6.303\pm 0.007$  \phn} & {\tiny$-1.011\pm 0.002  $  \phn} & {\tiny$-0.127\pm 0.007  $  \phn} & {\tiny$-0.058\pm 0.015  $  \phn} &  {\tiny} & {\tiny$6.84   $ } & {\tiny$350 \phn$ } & {\tiny$26.45^{+0.55 \phn}_{-2.45} $} \\
{\tiny BSDL349} & {\tiny$-5.910\pm 0.010$  \phn} & {\tiny$-0.841\pm 0.003  $  \phn} & {\tiny$-0.152\pm 0.010  $  \phn} & {\tiny$-0.062\pm 0.015  $  \phn} &  {\tiny} & {\tiny$7.08   $ } & {\tiny$350 \phn$ } & {\tiny$16.65^{+0.15 \phn}_{-1.35} $} \\
{\tiny BSDL34} & {\tiny$-5.936\pm 0.010$  \phn} & {\tiny$-0.835\pm 0.004  $  \phn} & {\tiny$-0.123\pm 0.010  $  \phn} & {\tiny$-0.038\pm 0.023  $  \phn} &  {\tiny} & {\tiny$7.11   $ } & {\tiny$350 \phn$ } & {\tiny$15.96^{+0.04 \phn}_{-0.96} $} \\
{\tiny BSDL2487} & {\tiny$-5.497\pm 0.023$  \phn} & {\tiny$-0.831\pm 0.008  $  \phn} & {\tiny$-0.106\pm 0.023  $  \phn} & {\tiny$-0.016\pm 0.043  $  \phn} &  {\tiny} & {\tiny$7.20   $ } & {\tiny$350 \phn$ } & {\tiny$13.87^{+0.03 \phn}_{-0.77} $} \\
{\tiny HS59.KMHK253} & {\tiny$-6.843\pm 0.005$  \phn} & {\tiny$0.597\pm 0.004  $  \phn} & {\tiny$0.845\pm 0.006  $  \phn} & {\tiny$0.417\pm 0.008  $  \phn} &  {\tiny} & {\tiny$7.20   $ } & {\tiny$350 \phn$ } & {\tiny$14.40^{+0.10 \phn}_{-0.60} $} \\
{\tiny BSDL1834} & {\tiny$-7.016\pm 0.005$  \phn} & {\tiny$-1.040\pm 0.002  $  \phn} & {\tiny$-0.072\pm 0.005  $  \phn} & {\tiny$0.067\pm 0.010  $  \phn} &  {\tiny} & {\tiny$6.69   $ } & {\tiny$400 \phn$ } & {\tiny$38.32^{+6.68 \phn}_{-1.32} $} \\
{\tiny BSDL2614} & {\tiny$-5.618\pm 0.012$  \phn} & {\tiny$-0.988\pm 0.005  $  \phn} & {\tiny$-0.176\pm 0.012  $  \phn} & {\tiny$-0.136\pm 0.028  $  \phn} &  {\tiny} & {\tiny$7.00   $ } & {\tiny$400 \phn$ } & {\tiny$19.11^{+0.14 \phn}_{-3.11} $} \\
{\tiny SL563} & {\tiny$-6.113\pm 0.012$  \phn} & {\tiny$-0.810\pm 0.004  $  \phn} & {\tiny$-0.137\pm 0.012  $  \phn} & {\tiny$-0.047\pm 0.024  $  \phn} &  {\tiny} & {\tiny$7.09   $ } & {\tiny$400 \phn$ } & {\tiny$16.50^{+0.10 \phn}_{-0.80} $} \\
{\tiny BSDL2725} & {\tiny$-6.354\pm 0.006$  \phn} & {\tiny$-0.955\pm 0.002  $  \phn} & {\tiny$-0.087\pm 0.006  $  \phn} & {\tiny$-0.069\pm 0.013  $  \phn} &  {\tiny} & {\tiny$6.96   $ } & {\tiny$450 \phn$ } & {\tiny$20.39^{+0.36 \phn}_{-1.89} $} \\
{\tiny BSDL2720} & {\tiny$-6.230\pm 0.007$  \phn} & {\tiny$-0.883\pm 0.003  $  \phn} & {\tiny$-0.034\pm 0.007  $  \phn} & {\tiny$0.006\pm 0.015  $  \phn} &  {\tiny} & {\tiny$7.03   $ } & {\tiny$450 \phn$ } & {\tiny$18.24^{+0.06 \phn}_{-1.04} $} \\
{\tiny H88-266} & {\tiny$-7.077\pm 0.009$  \phn} & {\tiny$-1.055\pm 0.004  $  \phn} & {\tiny$-0.206\pm 0.009  $  \phn} & {\tiny$-0.064\pm 0.017  $  \phn} &  {\tiny} & {\tiny$6.67   $ } & {\tiny$500 \phn$ } & {\tiny$39.74^{+7.96 \phn}_{-1.44} $} \\
{\tiny HS245} & {\tiny$-8.496\pm 0.002$  \phn} & {\tiny$0.559\pm 0.003  $  \phn} & {\tiny$1.172\pm 0.003  $  \phn} & {\tiny$0.472\pm 0.004  $  \phn} &  {\tiny} & {\tiny$6.82   $ } & {\tiny$500 \phn$ } & {\tiny$28.23^{+2.07 \phn}_{-0.23} $} \\
{\tiny BSDL305} & {\tiny$-7.801\pm 0.004$  \phn} & {\tiny$-0.941\pm 0.002  $  \phn} & {\tiny$-0.083\pm 0.004  $  \phn} & {\tiny$-0.015\pm 0.008  $  \phn} &  {\tiny} & {\tiny$6.63   $ } & {\tiny$600 \phn$ } & {\tiny$47.90^{+3.30 \phn}_{-2.40} $} \\
{\tiny BSDL2721} & {\tiny$-6.393\pm 0.008$  \phn} & {\tiny$-0.967\pm 0.003  $  \phn} & {\tiny$-0.050\pm 0.008  $  \phn} & {\tiny$-0.004\pm 0.017  $  \phn} &  {\tiny} & {\tiny$6.98   $ } & {\tiny$600 \phn$ } & {\tiny$19.62^{+0.28 \phn}_{-0.92} $} \\
{\tiny BSDL2583} & {\tiny$-6.081\pm 0.009$  \phn} & {\tiny$-0.960\pm 0.003  $  \phn} & {\tiny$-0.184\pm 0.009  $  \phn} & {\tiny$-0.106\pm 0.020  $  \phn} &  {\tiny} & {\tiny$6.98   $ } & {\tiny$700 \phn$ } & {\tiny$19.66^{+0.24 \phn}_{-0.96} $} \\
{\tiny HS74} & {\tiny$-6.560\pm 0.009$  \phn} & {\tiny$-0.844\pm 0.003  $  \phn} & {\tiny$-0.210\pm 0.009  $  \phn} & {\tiny$0.035\pm 0.020  $  \phn} &  {\tiny} & {\tiny$6.98   $ } & {\tiny$700 \phn$ } & {\tiny$19.83^{+0.07 \phn}_{-0.83} $} \\
{\tiny KMHK339} & {\tiny$-6.396\pm 0.008$  \phn} & {\tiny$-0.952\pm 0.003  $  \phn} & {\tiny$-0.158\pm 0.008  $  \phn} & {\tiny$-0.068\pm 0.020  $  \phn} &  {\tiny} & {\tiny$6.92   $ } & {\tiny$750 \phn$ } & {\tiny$22.56^{+0.04 \phn}_{-2.16} $} \\
{\tiny NGC2102.SL665} & {\tiny$-7.509\pm 0.004$  \phn} & {\tiny$-0.879\pm 0.002  $  \phn} & {\tiny$-0.118\pm 0.004  $  \phn} & {\tiny$-0.080\pm 0.010  $  \phn} &  {\tiny} & {\tiny$6.76   $ } & {\tiny$800 \phn$ } & {\tiny$32.73^{+3.27 \phn}_{-1.83} $} 

\enddata
\tablenotetext{a}{{\scriptsize Popescu et al. 2012}}
\tablenotetext{b}{{\scriptsize This work}}
\end{deluxetable}

\end{document}